%% file: main.tex
\documentclass[sigconf]{acmart}

\AtBeginDocument{%
  }

\copyrightyear{2026}
\acmYear{2026}
\setcopyright{cc}
\setcctype{by}
\acmConference[CHI '26]{Proceedings of the 2026 CHI Conference on Human Factors in Computing Systems}{April 13--17, 2026}{Barcelona, Spain}
\acmBooktitle{Proceedings of the 2026 CHI Conference on Human Factors in Computing Systems (CHI '26), April 13--17, 2026, Barcelona, Spain}
\acmPrice{}
\acmDOI{10.1145/3772318.3791962}
\acmISBN{979-8-4007-2278-3/2026/04}

\usepackage{multirow}
\usepackage{colortbl}
\newcommand{\x}[1]{{\leavevmode\color{black}{#1}}}

\definecolor{c1}{HTML}{9796bb}
\definecolor{c2}{HTML}{00beb9}
\definecolor{c3}{HTML}{dfb0c7}
\definecolor{c4}{HTML}{add9a1}
\definecolor{c5}{HTML}{eac793}

\acmSubmissionID{7745}

\begin{document}

\title{\x{The Evolving Duet of Two Modalities: A Survey on} Integrating Text and Visualization for Data Communication}
\author{Xingyu Lan}
\authornote{Xingyu Lan is also a member of the Research Group of Computational and AI Communication at Institute for Global Communications and Integrated Media.}
\email{xingyulan96@gmail.com}
\orcid{0000-0001-7331-2433}
\affiliation{%
  \institution{Fudan University}
  \city{Shanghai}
  \country{China}
}

\author{Xi Li}
\email{xi.li0903@gmail.com}
\affiliation{%
  \institution{Ant Group}
  \city{Shanghai}
  \country{China}
}

\author{Yixing Zhang}
\email{yixingzhang23@m.fudan.edu.cn}
\affiliation{%
  \institution{Fudan University}
  \city{Shanghai}
  \country{China}
}

\author{Mengqin Cheng}
\email{mqcheng16@fudan.edu.cn}
\affiliation{%
  \institution{Fudan University}
  \city{Shanghai}
  \country{China}
}

\author{Jiazhe Wang}
\email{neoddish@outlook.com}
\affiliation{%
  \institution{Ant Group}
  \city{Shanghai}
  \country{China}
}

\author{Siming Chen}
\authornote{Siming Chen is the corresponding author.}
\email{simingchen3@gmail.com}
\orcid{xxx}
\affiliation{%
  \institution{School of Data Science \& Qingdao Research Institute, Fudan University}
  \city{Shanghai}
  \country{China}
}

\newcommand{\etal}{et~al.~} 
\newcommand{\ie}{i.e.,~}
\newcommand{\eg}{e.g.,~}
\newcommand{\ncorpus}{220 }

\renewcommand{\shortauthors}{Lan et al.}

\renewcommand{\sectionautorefname}{Section}
\renewcommand{\subsectionautorefname}{Section}
\renewcommand{\subsubsectionautorefname}{Section}

\begin{abstract}
    Text plays a fundamental yet understudied role as a narrative device in data visualization. While existing research has extensively explored text as data input and interaction modality, its function in supporting storytelling and interpretation remains fragmented. To address this gap, this work presents a systematic review of 98 publications that provide insights into using text as narrative. We investigate how text can be utilized in visualization, analyze its functions and effects, and explore how it can be designed to facilitate data communication. Our synthesis identifies significant research gaps in this domain and proposes future directions to advance the integration of text and visualization, ultimately aiming to provide guidance for designing text that enhances narrative clarity and fosters engagement.
\end{abstract}

\ccsdesc[500]{Human-centered computing~Visualization design and evaluation methods}
\ccsdesc[500]{Human-centered computing~Visualization techniques}
\ccsdesc[500]{Human-centered computing~Interaction design}

\keywords{Narrative Visualization, Data Storytelling, Text Design, Text Visualization}


\maketitle

\input{Sections/01-intro.tex}

\input{Sections/02-related.tex}
\input{Sections/03-method}
\input{Sections/04-why}

\input{Sections/05-what}

\input{Sections/06-how}

\input{Sections/07-discussion}
\input{Sections/08-conclusion}


\begin{acks}
This work was supported by the National Natural Science Foundation of China (NSFC No.62402121, No.62472099), Ant Group, Shanghai Chenguang Program, and Research and Innovation Projects from the School of Journalism at Fudan University.
\end{acks}

\bibliographystyle{ACM-Reference-Format}
\bibliography{reference}


\end{document}

%% file: Sections/01-intro.tex
\section{Introduction}
\label{sec:intro}

Although the adage ``a picture is worth a thousand words'' is often used to highlight the power of visualization in conveying information, text remains an indispensable component of effective visualization. As Edward Tufte noted in his seminal book, \textit{The visual display of quantitative information}, ``Words and pictures belong together. Viewers need the help that
words can provide.''~\cite{tufte2001visual} 
Since ancient times, text has been a dominant communication medium in most societies, spanning the eras of handwriting, the printing press, and the internet. In recent years, the rise of large language models (LLMs) has further elevated the significance of text as a communication modality, prompting us visualization researchers to rethink and harness its enduring power.

\begin{figure*}[t!]
 \centering
 \includegraphics[width=\textwidth]{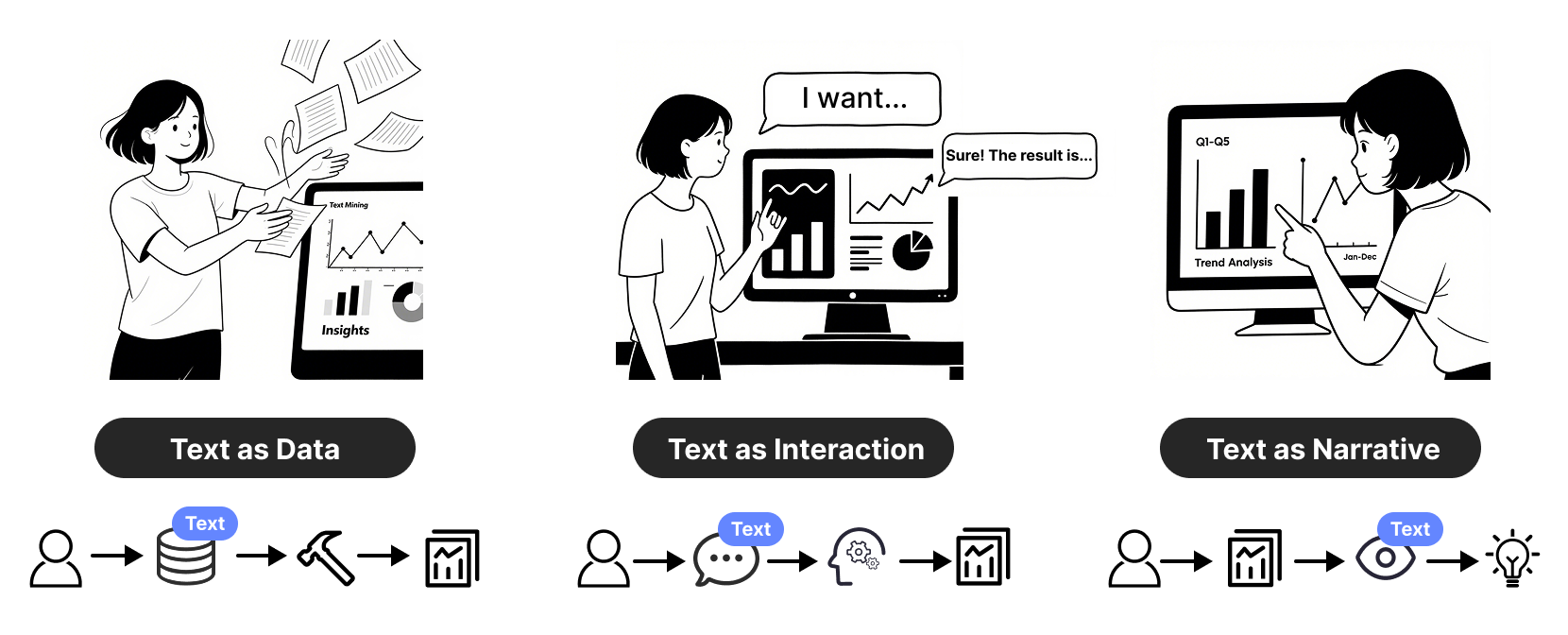}
 \caption{Three threads of research concerning text in the context of data visualization.}
 \label{fig:threads}
 \Description{The figure is organized into three vertically-aligned panels. The left panel, titled "Text as Data," depicts a girl inputting textual documents into a computer, representing text as a resource for computational processing. The central panel, titled "Text as Interaction," shows the girl engaging in a conversational dialogue with the computer, with speech bubbles containing "I want..." from the user and "Sure! The result is..." from the system, framing text as a medium for interactive communication. The right panel, titled "Text as Narrative," portrays the girl reading text accompanied by data visualizations on the screen, presenting text as a vehicle for explanation and storytelling.}
 \vspace{0em}
\end{figure*}

\x{However, research on text in visualization, while extensive, remains fragmented. The field often conflates the multifaceted roles of text—as data to be analyzed, a medium for interaction, and a vehicle for narrative explanation. This conceptual conflation impedes coherent scholarly discourse, as researchers may be speaking at cross-purposes about fundamentally different functionalities. Consequently, it may hinder the development of targeted theories and principles suited to each distinct textual role. To resolve this fragmentation and clarify the research gap, we categorize prior work into three main threads (see \autoref{fig:threads}):}

The first thread examines text as a type of data input (\ie \textbf{text as data}), focusing on developing techniques and systems for mining and visualizing textual input, such as social media posts~\cite{godwin2017typotweet,hu2016visualizing}, news~\cite{nguyen2018finanviz}, forum conversations~\cite{fu2018visforum}, and speech transcriptions~\cite{chandrasegaran2019talktraces}. 
Chronologically, this research theme emerged early and has manifested mainly as various visual analytics systems that apply state-of-the-art algorithms to text-mining tasks such as information retrieval, topic analysis, and sentiment analysis~\cite{li2022unified,cui2011textflow,kim2016topiclens}.
In addition, these studies have yielded a significant number of visualization designs tailored specifically for textual data~\cite{kucher2015text}, such as word clouds~\cite{wang2017edwordle,kulahcioglu2019paralinguistic}, SparkClouds~\cite{lee2010sparkclouds}, EventRiver~\cite{luo2010eventriver}, Textflow~\cite{cui2011textflow}, and TypoTweet~\cite{godwin2017typotweet}.
To date, this research thread has accumulated hundreds of papers and numerous valuable methods and insights~\cite{liu2018bridging}.

The second research thread treats text as an interaction channel (\ie \textbf{text as interaction}), a paradigm that has flourished alongside advances in natural language processing and generation. Representative work includes a growing body of chart-QA systems that allow users to obtain data analyses and visualizations by entering natural-language queries and to receive the answers in natural language as well~\cite{kim2020answering}. Example work includes systems and toolkit such as EVIZA~\cite{setlur2016eviza}, DataTone~\cite{gao2015datatone}, and NL4DV~\cite{narechania2020nl4dv}.
While early systems relied on structured phrasing to interpret user intent, recent advances can handle increasingly ambiguous, colloquial, and semantically rich input~\cite{setlur2019inferencing}. This improved interpretability has, in turn, spurred the development of many truly conversational visual analytics and visualization authoring systems~\cite{shen2022towards, kavaz2023chatbot, wang2022towards}.

The third thread, in contrast to the former two threads, examines text as a narrative device integrated with or accompanying data visualizations to aid in the comprehension and interpretation of data (\ie \textbf{text as narrative}). After all, as said by Tufte, ``it is nearly always helpful to write little messages on the plotting field to explain the data, to label outliers and interesting data points, to write equations and sometimes tables on the graphic itself, and to integrate the caption and legend into the design so that the eye is not required to dart back and forth between textual material and the graphic.''~\cite{tufte2001visual} 
In other word, only when graphical elements are effectively combined with text do they fully make sense to users. 
Regarding this, research about visualization design and data storytelling has provided many valuable insights. For instance, Borkin~\etal~\cite{borkin2015beyond} found that when viewing visualizations, viewers spend the most time on textual elements, especially the title. 
Kong~\etal~\cite{kong2018frames} found that people pay significant attention to the titles of visualizations, and more people perceive main messages from titles (even if they are slanted) rather than from the data visualizations.
Sultanum and Setlur~\cite{sultanum2024instruction} examined the interplay between text and visualizations in dashboards, highlighting ``the significance of elevating text as a first-class citizen in data visualization.''
Hearst~\cite{hearst2023show} also argued that ``language should be considered as co-equal with visualization when communicating information.''

However, compared to the other two threads, which have already been well-surveyed (\autoref{sec:related}), the knowledge of how to leverage the benefits of text as narrative remains significantly under-explored and unsynthesized. 
\x{While pioneering projects such as Explorable Explanations~\cite{victor} and Potluck~\cite{litt2022potluck} introduced the integration of text and graphics early on to overcome rigid and overly formal reading experiences, and while academia has accumulated a body of related work, to the best of our knowledge, most efforts remain scattered, and no peer-reviewed survey has yet addressed the third research thread.}
This gap may leave practitioners or future intelligent systems without clear guidance on how to intentionally design text to enhance narrative clarity, emphasize key insights, and foster engagement.
Besides, we would like to clarify that the three threads described above, while largely distinct, are not absolutely mutually exclusive. It is not uncommon for text to serve multiple roles simultaneously. For instance, a system may perform analytics on textual data while also employing textual narratives to present results vividly~\cite{sultanum2018more}; another might support the creation of text-rich data stories while also accepting natural language as input~\cite{fu2025dataweaver}. In this sense, a systematic investigation into text as narrative may also enhance the other two research threads by informing the development of more effective and user-centered systems.

Given the above motivations, in this work, we reviewed \x{98} papers that provide insights into using text as narrative in the context of data visualization. 
Specifically, we posed three main research questions:
\textbf{RQ1:} Why is it important to integrate text as narrative with visualization?
\textbf{RQ2:} What are the manifestations of text as a narrative device and its combination with visualization?
\textbf{RQ3:} How can text be designed to \x{facilitate explanatory data analysis and communication}?
By analyzing this body of literature, we discuss research gaps within each analyzed dimension and propose future research opportunities.

%% file: Sections/02-related.tex
\section{Related Surveys}
\label{sec:related}

The interplay between text and visualization has been explored from multiple angles, as outlined in the introduction. Several survey papers have already systematically reviewed the literature for the first two threads. This section reviews these existing surveys to clearly position our work and highlight the gap it aims to fill.

\textbf{Surveys on Text as Data.}
The first thread, which treats text as a data source to be mined and visualized, has been thoroughly reviewed. Specifically, Liu et al.~\cite{liu2018bridging} provided a task-driven survey that bridges the fields of text mining and visualization. By reviewing a substantial body of 263 papers, their work offers a systematic framework for visual text analytics, categorizing existing visualization and mining techniques according to analytical tasks such as information retrieval, cluster/topic analysis, classification, and trend analysis. In addition to this, Kucher and Kerren~\cite{kucher2015text} contributed a comprehensive taxonomy and survey of 141 text visualization techniques, providing valuable insights into the novel encodings for textual data. 
Because this research direction is well established, there is even a survey of surveys on text visualization~\cite{alharbi2018sos}. By reviewing all the state-of-the-art papers on text visualization, Alharbi~\etal~\cite{alharbi2018sos} categorized the surveys into 5 groups: document-centered, user task analysis, cross-disciplinary, multi-faceted, and satellite-themed.
Collectively, these works offer a robust foundation for understanding how to analyze and visualize textual content.

\textbf{Surveys on Text as Interaction.}
The second thread, which leverages text (particularly natural language) as an interaction modality, has also been the subject of recent surveys. 
For example, Shen et al.~\cite{shen2022towards} conducted a survey on natural language interfaces (NLI) for data visualization (which they called V-NLI). They systematically analyzed 57 papers about V-NLI, outlining the state-of-the-art techniques that enable users to query and generate visualizations using natural language.
Kavaz~\etal~\cite{kavaz2023chatbot} presented a scoping review about how chatbot-style NLIs are being coupled with data visualization systems, spotlighting current limits (\eg simple charts, rigid layouts, shallow guidance) and charting future avenues such as complex-data queries, AR/VR integration and smarter visual mapping.
Wang~\etal~\cite{wang2022towards} focused on natural language-based visualization authoring, proposing a NLI pipeline by introducing a structured representation of users' authoring utterances (\eg operations, objects, parameters). 
The above works meticulously cataloged the technical advances, interaction paradigms, and system designs, marking the rapid evolution of conversational visualization and intelligent authoring systems. 

\textbf{The Gap: A Survey on Text as Narrative.}
While the aforementioned surveys have extensively covered the technical underpinnings of using text as data and interaction, a significant gap remains. There is, to the best of our knowledge, no dedicated \x{peer-reviewed} survey that systematically investigates the use of text as a narrative device integrated within and alongside visualizations to aid comprehension and storytelling. 

The most relevant surveys to our work are those on data storytelling and visualization design. However, these prior studies generally adopt one of two approaches.
The first approach surveys the technical pipelines and tool ecosystems for data storytelling. For instance, Chen et al.~\cite{chen2023does} reviewed 105 papers and tools to characterize how automation can be engaged in the creation of narrative visualization. They categorized systems by their level of intelligence and automation techniques rather than delving deeply into the design and use of specific communicative elements. Similarly, He et al.~\cite{he2025leveraging} reviewed 66 papers to show how foundation models can be woven into the pipeline of narrative visualization, mapping their roles across four stages (\eg analysis, narration) and eight concrete tasks (\eg insight extraction, authoring). However, these surveys did not systematically examine the communicative function and design of textual components.
The second approach examines a wide variety of visualization forms, genres, and their constituent features to map the design space of information visualization. For example, Tong et al.~\cite{tong2018storytelling} provided a broad survey of existing techniques of data storytelling, while Zhao and Elmqvist~\cite{zhao2023stories} extended this effort with an expanded review of storytelling forms and components.
More recently, Rahman et al.~\cite{rahman2025annoation} conducted a dedicated survey on the usage of annotations in visualization. However, their review encompassed a wide range of graphical marks and highlighting effects, lacking a dedicated and focused investigation into textual elements specifically.

\x{Apart from the above surveys, we have also noted a latest paper by Stokes~\etal~\cite{stokes2025analysis} published recently that is very relevant to our work. The authors sampled 120 real-world visualizations, annotating 804 textual elements. Through open coding and factor analysis, they identified 10 common text functions (\eg present metadata, summarize values) and 4 high-level design patterns (\eg annotation-centric design, narrative framing).
Our work and theirs complement each other in several key aspects. First, while their study examines industry practices to reveal current uses of text in visualization, our work synthesizes academic literature to capture the state of the art. We believe that advanced design ideas from academia, which may not yet be widely adopted in the industry, are likely to provide valuable future directions for both researchers and practitioners.
Second, the analysis by Stokes et al. is limited to static images, while our review adopts a broader scope by including various dynamic web visualizations and interactive systems. This allows us to examine textual functions in richer contexts and tasks beyond static usage.
Last, Stokes et al.'s analysis employs a descriptive taxonomy to characterize existing text roles, while we construct a three-stage framework that incorporates the \textit{why}, \textit{what}, and \textit{how} of textual design. This allows us to synthesize a large amount of valuable arguments and empirical findings from academia (\autoref{sec:why}), as well as organize more than 30 actionable design techniques around core narrative tasks (\autoref{sec:space}). }

\x{In summary, our work complements prior research in both data sources and analytical focuses by offering a research-led perspective on text design for data communication.} By systematically reviewing the literature on text as a narrative device, we aim to consolidate currently scattered knowledge and establish a foundation for its effective design. Furthermore, this survey also serves to complete the missing segment of the three research threads concerning text and visualization.

%% file: Sections/03-method.tex
\section{Methodology}

In this section, we describe how we curated a corpus of relevant literature and analyzed it.

\subsection{Paper Search and Corpus Collection}

We tried two search strategies to identify relevant publications efficiently and effectively.
Our initial approach involved conducting keyword searches on multidisciplinary academic databases such as Google Scholar and Scopus, using queries requiring the presence of terms like ``text'' AND ``visualization'', ``data visualization'', or ``narrative visualization'' in paper titles, keywords, or abstracts. However, this method returned an unmanageably large number of results (tens of thousands), primarily due to significant impact arising from the rapid growth of NLP-related research—such as text mining and model training—that often incidentally included the target keywords. Even when narrowing the search to computer science-oriented databases like IEEE Xplore and ACM Digital Library, the result sets remained impractically large and chaotic.

Therefore, following previous surveys~\cite{shen2022towards,wang2021survey}, we refined our strategy to focus on specific venues in visualization. We systematically searched IEEE Xplore, ACM Digital Library, and the Eurographics Digital Library for publications from major visualization conferences and journals, including IEEE VIS, TVCG, PacificVis, EuroVis, Computer Graphics Forum, and CG\&A, as well as human-computer interaction venues including ACM CHI, UIST, IUI, and CSCW. Our keyword queries required that papers should include terms related to text (\eg text OR annotation OR title OR caption OR description) in combination with terms related to visualization (\eg chart OR graph OR diagram OR visualization). We also employed wildcards (e.g., visuali?z*) to account for spelling variations (e.g., visualize/visualise) and different word forms (e.g., visualization, visualizing).

This tailored approach yielded substantially more focused and precise results. Ultimately, we retrieved 769, 276, and 212 papers from the IEEE, ACM, and Eurographics libraries, respectively, resulting in a total of more than 1000 candidate papers. Next, we conducted a first round of manual screening based on titles and abstracts to exclude publications that were clearly unrelated to our research. For example, there were still many papers that focused primarily on text mining or purely textual analysis without any visualization component, which fall outside the scope of our survey (i.e., it should be at least research about visualization).
Subsequently, the remaining papers underwent a full-text review to identify those that genuinely contributed insights regarding text as narrative in the context of data visualization. This phase was carried out independently by two authors. 

Although this review process was largely straightforward, we discussed and iterated on two tricky issues.
Firstly, as noted in \autoref{sec:intro}, text can sometimes undertake compound roles. For instance, we came across some visualization systems that also include substantive contributions related to textual narrative (\eg wrapping analytical results into engaging text-visualization integrations such as data comics~\cite{zhao2021chartstory}). We therefore established a consensus-based inclusion criterion: systems or application papers were retained if they also contains contributions related to text as narrative.
In other words, we retained: 

\begin{itemize}
  \item papers that explicitly examined text as a narrative element (e.g., titles, captions, stories) in the context of data visualization (Criteria 1)
  \item papers that contain contributions concerning text as narrative, although the major contribution may be developing technologies and applications (Criteria 2)
\end{itemize}


Secondly, although our keyword search tried to cover a variety of common terms, some textual elements might be referred to using unconventional names, or might not be emphasized in the papers' metadata. To mitigate potential omissions, we manually supplemented our search with two methods: (i) consulting related surveys introduced in \autoref{sec:related} to view their paper lists and identify any missing qualified papers; (ii) using the snowballing method to check the related work cited by the included papers to identify any missing qualified papers~\cite{wu2021ai4vis}.

\begin{figure}[h]
 \centering
 \includegraphics[width=\columnwidth]{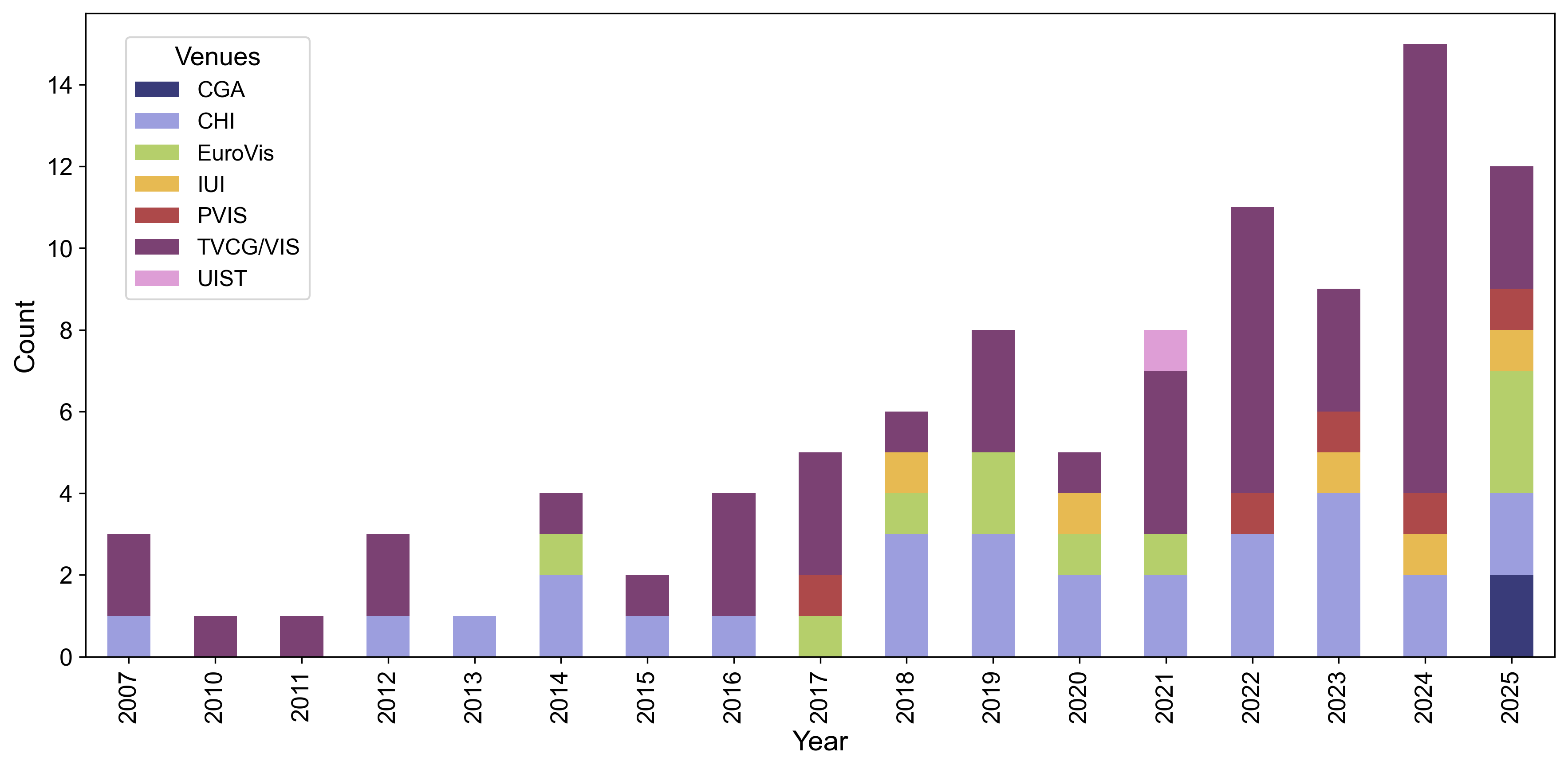}
 \caption{\x{Distribution of the 98 papers in our corpus.} Paper collection ended in August 2025.}
 \label{fig:corpus}
 \Description{This stacked bar chart illustrates the distribution of academic publications in our corpus from 2007 to 2025. The x-axis represents the year, ranging from 2007 to 2025, while the y-axis indicates the publication count from 0 to 14. Each bar corresponds to a specific year and is composed of stacked segments representing different venues: CGA, CHI, EuroVis, IUI, PVIS, TVCG/VIS, and UIST, each distinguished by color. The chart reveals a clear upward trend, beginning with approximately 3 publications in 2007 and rising to a peak of 15 in 2024. Throughout this period, TVCG/VIS and CHI appear as the dominant contributing venues.}
 \vspace{0em}
\end{figure}

Through this process, we ultimately assembled a final corpus of \x{98} qualified papers. In terms of publication venues, the largest share appeared in TVCG/VIS, followed by CHI, EuroVis, and PacificVis. Over time, publication counts demonstrate steady growth, with a notable acceleration in recent years (see \autoref{fig:corpus} for a summary), reflecting a significant surge of research interest in textual elements following the maturation of large language models. Within our corpus, 15 relevant papers were published in 2024 alone. Furthermore, as of our final search date in August 2025, 12 papers on this topic have already been published within 2025.

Of the \x{98} papers, \x{72} explicitly studied text as a narrative element (\ie meeting Criterion 1); typical examples are studies that investigated textual elements such as titles and descriptions in visualization design. The remaining 26 papers, 12 on text-data mining and 14 on natural-language interfaces for visualization, contributed text-as-narrative insights as a by-product (\ie meeting Criterion 2). 
For example, Medstory~\cite{sultanum2018more} is a visual-analytics system developed to help explore clinical reviews. While its primary focus is text mining and analysis, the authors highlighted that physicians need rich contextual information to obtain a credible medical narrative. The paper thus also proposed methods to enhance data communication through text (\ie combing \textit{text as data} and \textit{text as narrative}). As another example, DataParticles~\cite{cao2023dataparticles} can extract data entities from free-form text and generate animated unit visualizations; because the authoring system supports public sharing, it also includes rich narrative-oriented features such as text highlighting and explanatory sentences (\ie combining \textit{text as interaction} and \textit{text as narrative}). 
Such works illustrate the growing cross-fertilization between visual analytics and visual communication.



\subsection{Coding Process}

We coded the papers following the why-what-how axes, which is a well-established survey framework in visualization research~\cite{wu2021ai4vis,rahman2025annoation}. \x{For the \textit{why} dimension, we focused on the motivations for integrating text and visualization for data communication.
Specifically, following the methodology of prior work~\cite{lan2023affective,lan2025more}, we first manually read the full text of each paper and then marked argumentative sentences establishing the rationale for integrating text and visualization through claims of its necessity, importance, or urgency. Then, we} performed bottom-up thematic distillation of their underlying rationales.
\x{During this process, three major themes emerged: some papers (\eg ~\cite{parnow2015micro,latif2021kori}) directly cited established theories to support their claims (coded as \textit{established theories}); some drew upon domain requirements (e.g. journalism~\cite{chen2023calliope,fu2025dataweaver}) to articulate their motivations; while others conducted user studies themselves (e.g., \cite{kong2019trust,zhi2019linking}) to demonstrate the importance of integrating text and visualization (coded as \textit{empirical study}). For papers that contribute original user studies, we further documented the conditions they examined and text-related findings.}

For the \textit{what} dimension, we concentrated on the manifestations of text and the specific ways it is combined with visualization. \x{We first extracted the original naming of textual elements (\eg ``title'', ``article'') examined by the papers, and then grouped synonymous terms to summarize common textual forms. The main challenge we encountered was the inconsistent wording used by different authors. For instance, while some papers employ the term ``description'' to refer to short explanatory text accompanying visualizations (typically ranging from a sentence to a paragraph), others use ``caption'' to denote similar short descriptions~\cite{zhao2021chartstory, kim2021towards}. Given the functional similarity between these text forms and the absence of a clear boundary between ``caption'' and ``description'', we uniformly refer to all such short explanatory text as \textit{description}. Additionally, we coded the integration modes of text and visualization by drawing on spectrum-based frameworks from prior work (\eg the categorization of ``chart in text'', ``separate'', and ``text in chart'' in~\cite{masson2023charagraph} and ``text-centric'', ``visualization-centric'', and ``non-centric'' in~\cite{cai2024linking}).}

For the \textit{how} dimension, we examined low-level text design methods to identify concrete methods of using text to aid data communication. 
Following previous design-space studies, we generated open codes to describe design techniques, then classified and structured them into a taxonomy.
However, because our focus is on how text serves data communication and narrative rather than on mining or generation tasks, arranging codes along a technical pipeline—the conventional practice in visualization surveys—would deviate from our core research questions and obscure our contribution. After iterative discussions, we therefore mapped the identified design techniques to distinct narrative tasks, drawing on prior literature~\cite{shi2021communicating}. This ensures that the findings of this section are tightly aligned with the central theme of our paper. \x{As a result, five narrative tasks were identified: \textit{Explain}, \textit{Emphasize}, \textit{Couple}, \textit{Adapt}, and \textit{Verify}. This categorization underwent careful iteration. For example, our initial framework distinguished between \textit{Summarize} and \textit{Explain}. During coding, however, we observed that data summaries often function inherently as explanations. Consider a trend line labeled ``steady growth'': it can be viewed both as a summary of the data and as an explanation of the overall pattern—each providing interpretation beyond the raw values. As the designer’s objective in both cases is to make the data intelligible, we merged them into the \textit{Explain} task.}


The codes of all collected papers can be found in the supplemental file of this work.


%% file: Sections/04-why.tex
\section{Why Integrating Text and Visualization}
\label{sec:why}

In visualization research, the value of text appears paradoxical. Although its importance seems intuitive, and we rarely encounter a visualization devoid of text, text is nevertheless often cast in a supporting or secondary role being \textit{added} to graphical elements~\cite{stokes2024give}. Addressing this tension, we noticed that most papers (\x{83 out of 98}) in our corpus have more or less offered responses and arguments. In this section, we report these arguments as three categories: (i) those driven by real-world domain and user needs, (ii) those grounded in established theories, and (iii) those based on empirical studies.

\subsection{Domain and User Needs}
\label{ssec:domain_needs}

More than half of the papers (\x{N = 62, 63.3\%}) in our corpus have positioned narrative text as indispensable for meeting certain domain or user requirements.
For example, 12 papers\x{~\cite{hullman2013contextifier,gao2014newsviews,boy2015storytelling,fulda2015timelinecurator,kong2018frames,riederer2018put,metoyer2018coupling,kong2019trust,latif2021deeper,chen2023calliope,fu2025dataweaver,lecardonnel2025genqa}} are explicitly grounded in the needs of the \textbf{journalism industry}, demonstrating that textual elements help explain data insights and enable non-experts to understand visualizations more effectively. Latif~\etal~\cite{latif2021deeper} argued that ``data-driven stories presented in online articles combine the expressive power of visualizations with a textual narrative. In these stories, visualizations provide an overview of the data while the accompanying text highlights insights and blends in the backdrop of the story.''
Nine papers\x{~\cite{jung2021communicating,lundgard2021accessible,alam2023seechart,seo2024maidr,moured2024chart4blind,xu2024graphs,li2024altgeoviz,mcnutt2025accessible,lecardonnel2025genqa}} focus on the needs of \textbf{visually impaired users}, arguing that text is a critical channel for making visualizations accessible. Through interviews with this group, Mcnutt~\etal~\cite{mcnutt2025accessible} found that ``the text descriptions gave participants a `decent sense as to like what might look like' and that they were informative and appropriate in length.'' 
Seven papers\x{~\cite{chen2010click2annotate,elias2012annotating,zheng2022telling,bartram2021untidy,wang2023slide4n,li2023notable,zhao2021chartstory}} claim to serve the needs of \textbf{data scientists}. For instance, researchers~\cite{elias2012annotating,li2023notable,zheng2022telling} have highlighted the necessity of using textual elements to ``externalize'' analytical insights and communicate findings to others. Bartram~\etal~\cite{bartram2021untidy} conducted interviews with data workers, finding that they indeed frequently add marginalia—such as comments and various annotations—when working with spreadsheets.
Six papers\x{~\cite{zou2025gistvis,masson2023charagraph,wang2020data,latif2018vis,beck2017word,dragicevic2019increasing}} focus on the needs of \textbf{research communication}. Wang~\etal~\cite{wang2020data} argued that ``effective communication in academic research is crucial, as it allows readers to assess the rigor of scientific methods and to build trust in scientific results. '' However, as stated by Dragicevic~\etal~\cite{dragicevic2019increasing}, ``the recent replication crisis in psychology and other disciplines has dealt a blow to the credibility of human-subject research and prompted a movement of methodological reform'', thus calling for more integrated and interactive presentation of text and visualization in scientific reporting.
Two papers\x{~\cite{sultanum2018more,sultanum2018doccurate} by Sultanum~\etal} focused on the \textbf{medical industry}, claiming that clinicians need more textual context, not less. They conducted a series of studies in this area. They interviewed physicians and found that textual notes are essential communicative devices for them.

Apart from these, \x{17} papers\x{~\cite{kong2014extracting,satyanarayan2014authoring,goffin2014exploring,kim2016generating,ren2017chartaccent,badam2018elastic,goffin2020interaction,sultanum2021leveraging,conlen2021idyll,lalle2021gaze,zheng2022evaluating,latif2021kori,sultanum2023datatales,shen2023data,poehls2025either,cai2024linking,sultanum2024instruction}} mentioned \textbf{mixed domain needs}. For example, Kong~\etal~\cite{kong2014extracting} claimed that ``news articles, reports, blog posts and academic papers often include graphical charts that serve to visually reinforce arguments presented in the text.'' Shen~\etal~\cite{shen2023data} demonstrated that ``visualization and corresponding descriptions often work together for data storytelling. Combining data visualization and narratives, data videos have become popular among practitioners as a visual storytelling form in fields such as journalism, marketing, and education.''
Another five papers\x{~\cite{steinberger2011context,kandogan2012just,bryan2016temporal,sauve2022put,novotny2024evaluating}} highlighted \textbf{technological and hardware constraints}. For example, Steinberger~\etal~\cite{steinberger2011context} argued that it is essential to design explicit links between text and visualization because they can connect elements that fall outside the small active visual field. This is especially important given the trend toward large, high-resolution displays.
Novotny~\etal~\cite{novotny2024evaluating} argued that ``displaying text legibly yet space-efficiently is a challenging problem in immersive displays.''


\subsection{Theoretical Underpinnings}

\x{When arguing for the integration of text and visualization}, 16 papers have cited established theories from various fields.
Among these, \x{9} papers\x{~\cite{steinberger2011context,parnow2015micro,beck2017word,kong2018frames,latif2021kori,latif2018exploring,barral2020understanding,lalle2021gaze,poehls2025either}} cited theories from psychology \x{and 5 of them~\cite{latif2021kori,latif2018exploring,beck2017word,lalle2021gaze,barral2020understanding} referred to the \textbf{split-attention effect}, which posits that ``learners are often forced to split their attention between and mentally integrate disparate sources of information (e.g., text and diagrams) before the instructional material can be rendered intelligible,'' so that ``physical integration (e.g., combining text and diagrams) may reduce cognitive load and so facilitate learning''~\cite{chandler1992split}. 
Because this theory aligns closely with many visualization-design scenarios, researchers have used it to argue that traditional text–image separation fails to exploit the full value of text and highlight the necessity of enhancing text-visualization linkage in explanatory visualizations such as data-driven articles~\cite{lalle2021gaze} and data-rich documents~\cite{latif2021kori}.}
Another referenced theory is \textbf{Gestalt principles}\x{~\cite{steinberger2011context}}, which emphasize \x{the grouping effect when humans perceive visual elements}. 
\x{Therefore, Steinberger~\etal~\cite{steinberger2011context}} argued that by aligning text with visual elements in accordance with Gestalt principles \x{(\eg proximity, similarity, closure, and continuity)}, designers can create more intuitive and coherent visual narratives that facilitate quicker and more accurate understanding.
In addition, some researchers draw on psychological studies to point out that improper use of text can also be detrimental.
For example, Parnow~\etal~\cite{parnow2015micro} cited research on \textbf{cognitive load}, stating that ``an interwoven presentation can help the reader to better comprehend the information.'' 
Kong~\etal~\cite{kong2018frames}, motivated by previous work on \textbf{cognitive biases}, chose to examine the slanted titles of data visualization.

\x{Three papers referred to theories from learning science~\cite{zhi2019linking,arunkumar2025lost,stokes2024give}.}
Zhi~\etal~\cite{zhi2019linking} cited Mayer’s \textbf{multimedia learning theory}, which states that ``learning can be seen as having three kinds of cognitive demands: essential processing, incidental processing, and representational holding'', arguing that ``the associated representational holding can be potentially reduced and essential processing can be in- creased if readers can easily link the text and pictures in multimedia content.'' Also drawing upon this theory, Arunkumar~\etal~\cite{arunkumar2025lost} argued that ``people learn more effectively when words and images are combined, rather than used alone.''
Stokes~\etal~\cite{stokes2024give} cited research about \textbf{fluent reading} to highlight the importance of appropriate text design.
Poehls~\etal~\cite{poehls2025either}, based on the theory about \textbf{dual-coding learning styles}, examined whether text-rich data articles would outperform data videos.

One paper drew on the \textbf{information-foraging theory}~\cite{willett2007scented} from behavioral ecology, arguing that ``improving information scent through better proximal cues lowers the cost structure of information foraging and improves information access.''
Another study on geovisualization\cite{kim2016generating} borrowed the geography theory of ``\textbf{landmarks}'', noting that people’s mental maps of space and distance are metrically distorted; inserting familiar places and scales into textual descriptions will make data communication easier.
Finally, one work cited \textbf{narrative theory}~\cite{eccles2008stories}, which suggests that ``people are natural storytellers who intuitively evaluate a story for consistency, detail, and structure'' to demonstrate the necessity of embedding textual annotations and other storytelling mechanisms in visual analytics.

Together, these theoretical foundations justify the use of text in visualizations, as well as the necessity of designing text to enhance overall communication effectiveness.

\subsection{Empirical Findings}

Apart from citing established theories, we also observed \x{23} papers that have conducted empirical studies themselves to assess the effect of text in the context of data communication (see \autoref{tab:empirical} for summarization). Notably, we observed that a considerate number of studies have used controlled comparisons to examine the impact of text in visualizations. These studies help quantify the effect of text more precisely, especially the exact value of utilizing text or combining text and visualization.

\input{Tables/empirical}

Most of these studies concentrate on the effect of text on how people read and understand visualizations, especially their \textbf{comprehension \& understanding} and \textbf{memory}.
Kong~\etal~\cite{kong2018frames} and Stokes~\etal~\cite{stokes2023role}, for example, found that the content of textual elements strongly affect how people perceive the message of a visualization.
Kim~\etal~\cite{kim2021towards} found that the phrasing of captions influences users' takeaway messages.
Ajani et al.~\cite{ajani2021declutter} compared cluttered visualizations against decluttered versions and decluttered versions plus focused annotations. They found that adding meaningful textual annotations improved data comprehension effectively, and that retaining key textual labels while removing redundant grid and legend text optimized the balance between clarity and context.

Some have assessed \textbf{subjective engagement}, such as \textbf{preference} and \textbf{enjoyment}.
Stokes and Hearst~\cite{stokes2022more} argued that the role of textual annotations is underestimated and lacks systematic study. Using line charts as examples, they created multiple variants such as chart-only to chart+title, chart+title+annotation, and text-only. Through crowdsourcing, they found that people generally prefer charts with more textual annotations. 
Although annotations may appear cluttered sometimes, participants favored versions that provided richer context and information. 
Wang~\etal~\cite{wang2019comparing} conducted two user studies comparing three text–visualization integration forms, \ie illustrated text, infographic, and data comic, for user engagement and recall. They found that forms in which text and visuals are tightly integrated, such as comics, yielded significantly higher immediate recall and self-reported enjoyment.
Mckenna et al.~\cite{mckenna2017visual} compared user engagement across four data story designs: text only, text with static visualizations, text with visualizations and stepper transitions, and text with visualizations and scroller transitions. Their results showed that user engagement increased significantly in conditions where text was deeply integrated with visualizations.
A small number of papers have examined other factors, such as \textbf{trust}, \textbf{attitude}, \textbf{gaze behavior}, information-retrieval \textbf{efficiency}, and \textbf{exploratory behaviors}.
For example, in an eye-tracking study, Borkin~\etal~\cite{borkin2015beyond} found that people allocate the most time to the title when viewing a data visualization. Zheng~\etal~\cite{zheng2022evaluating} found that linking text and visualization in data-driven articles helps enhance perceived credibility and people's ability to detect misinformation.

In terms of the effects of text (summarized in the ``effects'' column in \autoref{tab:empirical}, the majority of studies have identified significant advantages of using text or integrating text and visualization, especially regarding comprehension and memory.
However, a few studies yielded different results.
As one case, Boy~\etal~\cite{boy2015storytelling} found that ``augmenting an exploratory visualization with initial narrative visualization techniques and storytelling does not increase user engagement in exploration.''
Ottley~\etal~\cite{ottley2019curious} performed an eye-tracking study and found that even when text and visualization are presented together, ``users do not integrate information well across the two representation types.''
Poehls~\etal~\cite{poehls2025either} found that although interactive articles are equally effective in terms of information retention compared to data videos, it required about 30\% more time for the median user to absorb the same amount of information.
However, most studies summarized in \autoref{tab:empirical} have provided substantial empirical evidence supporting the necessity of textual narratives in data visualization.

%% file: Tables/empirical.tex

\makeatletter
\def\arraystretch{1.4}      
\makeatother

\begin{table*}[htbp]
    \fontsize{8}{8.7}\selectfont
    \centering
    \begin{tabular}{p{2.2cm}|p{2.5cm}p{11cm}c}
    \toprule
    \rowcolor[HTML]{E4E9FF} 
    \textbf{Measurement} & \textbf{Literature} & \textbf{Tested conditions} & \textbf{Effect} 
    \\ \hline 

    \multirow{12}{=}{\textbf{Comprehension \& Understanding}} & {Riederer et al.~\cite{riederer2018put}} & {data description with / without a perspective statement} & {\color{green!70!black}$\bullet$}  \\ 
    {} & {Kong et al.~\cite{kong2018frames}} & {supporting / non-supporting title for visualization} & {\color{green!70!black}$\bullet$} \\ 
    {} & {Wang et al.~\cite{wang2019comparing}} & {illustrated text \x{/} infographic \x{/} data comic} & {\color{green!70!black}$\bullet$}  \\ 
    {} & {Stokes et al.~\cite{stokes2022striking}} & {chart only \x{/} chart+title+annotation \x{/} chart+title+annotation+narrative \x{/} text only} & {\color{green!70!black}$\bullet$}  \\ 
    {} & {Fan et al.~\cite{fan2024understanding}} & {\x{3 map types * 4 semantic levels *  aligned/unaligned text-map detail levels}} & {\color{green!70!black}$\bullet$}  \\ 
    {} & {Zhi et al.~\cite{zhi2019linking}} & {vertical/slideshow layout * linking/not linking text and visualization} & {\color{gray!60}$\bullet$}  \\
    {} & {Poehls~\etal~\cite{poehls2025either}} & {interactive data-driven article \x{/} data video} & {\color{gray!60}$\bullet$}  \\ 
    {} & {Burns~\etal~\cite{burns2022invisible}} & {visualization only \x{/} metadata only \x{/} visualization+metadata} & {\color{gray!60}$\bullet$}  \\ 
    {} & {Kim~\etal~\cite{kim2021towards}} & {\x{caption that describes the prominent (1st, 2nd, or 3rd) chart feature / caption that describes non-prominent feature / no caption}} & {\color{gray!60}$\bullet$}  \\ 
    {} & {Ottley~\etal~\cite{ottley2019curious}} & {text only \x{/} visualization only \x{/} text+visualization} & {\color{orange!70!black}$\bullet$} \\ 
    {} & {\x{Lall{\'e}~\etal~\cite{lalle2021gaze}}} & {\x{gaze-driven highlighting of text-visualization linkages / no highlighting}} & {\color{orange!70!black}$\bullet$} \\ 
    \hline
    
    \multirow{8}{=}{\textbf{Memory}} & {Borkin et al.~\cite{borkin2015beyond}} & {view visualization and later recall it \x{(no controlled conditions)}} & {\color{green!70!black}$\bullet$}  \\ 
    {} & {Kong et al.~\cite{kong2018frames}} & {supporting/non-supporting title for visualization} & {\color{green!70!black}$\bullet$}  \\ 
    {} & {Kong et al.~\cite{kong2019trust}} & {\x{selective slant / miscued slant / contradictory slant in visualization titles}} & {\color{green!70!black}$\bullet$}  \\
    {} & {Riederer et al.~\cite{riederer2018put}} & {data description with / without a perspective statement} & {\color{green!70!black}$\bullet$} \\
    {} & {Wang et al.~\cite{wang2019comparing}} & {illustrated text \x{/} infographic \x{/} data comic} & {\color{green!70!black}$\bullet$}  \\
    {} & {Ajani et al.~\cite{ajani2021declutter}} & {cluttered \x{/} decluttered \x{/} decluttered+focus} & {\color{green!70!black}$\bullet$}  \\
    {} & {Arunkumar et al.~\cite{arunkumar2023image}} & {view visualization and later recall it \x{(no controlled conditions)}} & {\color{green!70!black}$\bullet$}  \\
    {} & {Zhi et al.~\cite{zhi2019linking}} & {vertical/slideshow layout * linking/not linking text and visualization} & {\color{gray!60}$\bullet$}  \\
    \hline

    \multirow{10}{=}{\textbf{Subjective Experience (\eg preference, enjoyment, perceived relevance)}} & {Mckenna et al.~\cite{mckenna2017visual}} & {text \x{/} text+static visualization \x{/} text+visualization+stepper \x{/} text+visualization+scroller} & {\color{green!70!black}$\bullet$}  \\ 
    {} & {Wang et al.~\cite{wang2019comparing}} & {illustrated text \x{/} infographic \x{/} data comic} & {\color{green!70!black}$\bullet$}  \\ 
    {} & {Zhi et al.~\cite{zhi2019linking}} & {vertical/slideshow layout * linking/not linking text and visualization} & {\color{green!70!black}$\bullet$}  \\
    {} & {Ajani et al.~\cite{ajani2021declutter}} & {cluttered \x{/} decluttered \x{/} decluttered+focus} & {\color{green!70!black}$\bullet$}  \\
    {} & {Stokes et al.~\cite{stokes2022more}} & {chart only \x{/} chart+title \x{/} chart+title+annotation \x{/} chart+title+2 annotations \x{/} chart+title+3\textasciitilde6 annotations \x{/} text only} & {\color{green!70!black}$\bullet$} \\
    {} & {Stokes et al.~\cite{stokes2022striking}} & {chart only \x{/} chart+title+annotation \x{/} chart+title+annotation+narrative \x{/} text only} & {\color{green!70!black}$\bullet$}  \\ 
    {} & {Burns~\etal~\cite{burns2022invisible}} & {visualization only \x{/} metadata only \x{/} visualization+metadata} & {\color{gray!60}$\bullet$} \\
    {} & {\x{Lall{\'e}~\etal~\cite{lalle2021gaze}}} & {\x{gaze-driven highlighting of text-visualization linkages / no highlighting}} & {\color{gray!60}$\bullet$} \\ 
    {} &  {\x{Burns~\etal~\cite{burns2022invisible}}} & {\x{visualization only \x{/} metadata only \x{/} visualization+metadata}} & {\color{orange!70!black}$\bullet$}  \\ 
    \hline

    \multirow{5}{=}{\textbf{Trust \& Perceived Bias}} &  {Kong et al.~\cite{kong2018frames}} & {supporting / non-supporting title for visualization} & {\color{green!70!black}$\bullet$} \\ 
    {} & {Kong et al.~\cite{kong2019trust}} & {\x{selective slant / miscued slant / contradictory slant in visualization titles}} & {\color{green!70!black}$\bullet$} \\
    {} & {Burns~\etal~\cite{burns2022invisible}} & {visualization only \x{/} metadata only \x{/} visualization+metadata} & {\color{green!70!black}$\bullet$}  \\ 
    {} &{Zheng~\etal~\cite{zheng2022evaluating}} & {\x{static illustrated text / explanatory annotation / interactive text-visualization linking in data-driven articles}} & {\color{green!70!black}$\bullet$} \\ 
    {} & {Stokes~\etal~\cite{stokes2023role}} & {\x{text as title/caption * 4 semantic levels (study 1); no-side / low-bias / high-bias wording (study 2)}} & {\color{green!70!black}$\bullet$} \\ 
    \hline

    \multirow{3}{=}{\textbf{Gaze Behavior}}  
    & {Borkin et al.~\cite{borkin2015beyond}} & {visualization viewing \x{(no controlled conditions)}} & {\color{green!70!black}$\bullet$} \\
    & {Barral et al.~\cite{barral2020understanding}} & {data-driven articles with / without text-visualization linkage} & {\color{green!70!black}$\bullet$} \\
    {} & {Ottley~\etal~\cite{ottley2019curious}} & {text only \x{/} visualization only \x{/} text+visualization} & {\color{orange!70!black}$\bullet$} \\ 
    \hline

    \multirow{3}{=}{\textbf{Efficiency}} &     
    {Zhi et al.~\cite{zhi2019linking}} & {vertical/slideshow layout * linking/not linking text and visualization} & {\color{gray!60}$\bullet$}  \\
    {} & {\x{Lall{\'e}~\etal~\cite{lalle2021gaze}}} & {\x{gaze-driven highlighting of text-visualization linkages / no highlighting}} & {\color{gray!60}$\bullet$} \\
    {} & {\x{Poehls~\etal~\cite{poehls2025either}}} & {\x{interactive data-driven article \x{/} data video}} & {\color{orange!70!black}$\bullet$} 
    \\ 
    \hline

    \multirow{1}{=}{\textbf{Prediction}} &     
    {Stokes~\etal~\cite{stokes2023role}} & {\x{text as title/caption * 4 semantic levels (study 1); no-side / low-bias / high-bias wording (study 2)}} & {\color{gray!60}$\bullet$} \\ 
    \hline


    \multirow{2}{=}{\textbf{Exploratory Behavior}} & 
    {Zhi et al.~\cite{zhi2019linking}} & {vertical/slideshow layout * linking/not linking text and visualization} & {\color{gray!60}$\bullet$}  \\
    {} & {Boy~\etal~\cite{boy2015storytelling}} & {visualization with / without an introductory narrative component} & {\color{orange!70!black}$\bullet$} \\ 
    \hline

    \multirow{2}{=}{\textbf{Attitude Change}} & {Kong~\etal~\cite{kong2018frames}} & {supporting / non-supporting title for visualization} & {\color{orange!70!black}$\bullet$} \\ 
    {} & {Liem et al.~\cite{liem2020structure}} & {\x{exploratory design / empathy design (personal story) / structure design}} & {\color{orange!70!black}$\bullet$} \\


    \bottomrule
    \end{tabular}
    \vspace{0em}
    \caption{Literature that empirically evaluated the effects of text as narrative in the context of data visualization. For the ``Effect'' column, {\color{green!70!black}$\bullet$} denotes positive effects concerning text, {\color{gray!60}$\bullet$} denotes neutral/mixed effects, and {\color{orange!70!black}$\bullet$} denotes negative effects.}
    \Description{This table summarizes empirical studies on text-as-narrative in data visualization, listing the measurement type, associated literature, tested experimental conditions, and the observed effect (positive, neutral, or negative). You can interact with the table directly.}
    \label{tab:empirical}
    \vspace{0em}
\end{table*}

%% file: Sections/05-what.tex
\section{What Are the Manifestations of Text?}
\label{sec:space}

Our analysis of the \textit{what} dimension has revealed rich manifestations of text and its diverse modes of integration with visualization, implying the broad application scope and design possibilities of text as narrative.

\begin{figure*}[t]
 \centering
 \includegraphics[width=\textwidth]{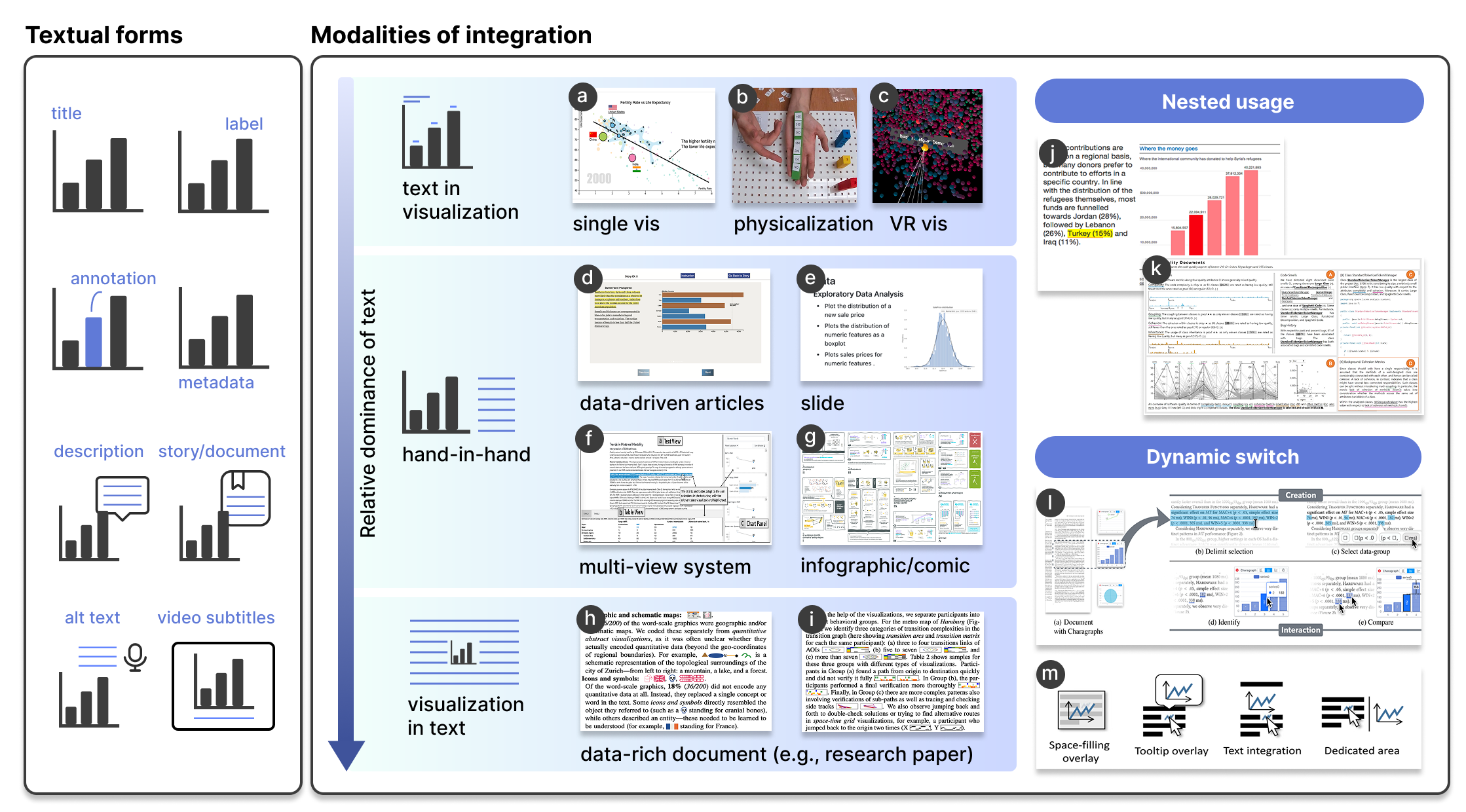}
 \caption{Textual forms (left) and their modalities of integration with visualization (right). \x{Examples of these modalities include: (a) single visualization~\cite{ren2017chartaccent}, (b) labeled data physicalization~\cite{sauve2022put}, (c) labeled VR visualization~\cite{novotny2024evaluating}, (d) data-driven article~\cite{zhi2019linking}, (e) slideshow~\cite{zheng2022evaluating}, (f) multi-view system~\cite{badam2018elastic}, (g) infographic/comic~\cite{wang2020data}, (h\&i) data-rich documents~\cite{beck2017eye,goffin2016exploratory}, (j) using text within a visualization as well as in parallel with the visualization~\cite{kong2014extracting}, (k) using separate panels to show text and visualizations while also embedding small inline charts within the text panel~\cite{mumtaz2019exploranative}, (l) Charagraph enables bi-directional interaction and switch between text and chart~\cite{masson2023charagraph}, (m) several assembly forms of text and image that can be flexibly switched proposed by Beck~\etal~\cite{beck2017word}.}}
 \label{fig:manifest}
 \Description{The figure is divided into two vertical sections. On the left, under the heading "Textual forms," a list of text types is shown with simple bar chart icons. These are: title, label, annotation, metadata, description, story/document, alt text, and video subtitles. On the right, under the heading "Modalities of integration," the section is organized into several horizontal bands. A large vertical axis labeled "Relative dominance of text" runs from top to bottom, indicating a spectrum from visualization-dominate to text-dominate. Top Band (Text in Visualization): Shows examples where text is embedded within visualizations. Examples include: (a) Single vis, (b) Physicalization, (c) VR vis. Middle Band (Hand-in-hand): Shows examples where text and visualization are presented side-by-side or integrated closely: (d) Data-driven articles, (e) Slide, (f) Multi-view system, (g) Infographic/Comic. Bottom Band (Visualization in Text): Shows examples where visualizations are embedded within a larger body of text, like a (h) Data-rich document. To the far right, two additional concepts are illustrated: Nested usage: Two examples showing how different integration modalities introduced above can be combined. Dynamic switch: Two examples showing how different integration modalities introduced above can be switched through interaction.}
 \vspace{0em}
\end{figure*}

\subsection{Textual Forms}

Text within data visualization is not a monolithic entity; it manifests across a spectrum of scales, each serving distinct communicative purposes. This spectrum ranges from concise, simple identifiers to extensive narratives that frame the entire visualization. Based on our analysis of the literature, we categorized these manifestations into eight major types. Below, we present them one by one.

\textit{Title} (N = 15) is a concise sentence or phrase that states the key finding or hook of a visualization, actively framing how the viewer should interpret the graphic that follows~\cite{kong2018frames,liu2023autotitle}.
\textit{Label}  (N = 6) is a textual component that displays data values and attributes. It consists of short strings—axis labels, legend entries, data-point identifiers—whose primary role is disambiguation and reference, allowing the viewer to decode the visual encodings. In some interactive systems, labels can also be implemented as textual tooltips~\cite{sultanum2024instruction}.
\textit{Metadata} (N = 3) is brief, compact text typically placed in footnotes or sub-headers to supply basic facts such as data sources, the time of the last update, or author information. Across different papers these textual forms are given inconsistent names, some called them ``additional information'', while others mixed them with ``captions'' or ``descriptions''. To avoid confusion with later categories, we refer to the work by Burns~\cite{burns2022invisible}, who defined metadata as six specific types of text (\eg data source, the transformations applied to the data, the people involved in its creation).
Next, textual annotation (N = 25), often deployed as call-outs (also termed additive cues~\cite{hullman2013contextifier}), highlights and explains specific data features, or point of interest (POI)~\cite{bryan2016temporal,li2024viscollage}, such as outliers, peaks, clusters. Unlike generic labels, annotations can supply deeper interpretive insight or additional contextual information.
However, one tricky point is that the term ``annotation'' is used inconsistently across papers. For example, Click2Annotate~\cite{chen2010click2annotate} called a paragraph placed separately below a visualization as an annotation, and many works treat purely visual effects, such as coloring or outlining a single bar in a bar chart, as annotations. In this sense, under the current research landscape, textual annotation is, strictly speaking, an important subset of annotation (Rahman~\etal~\cite{rahman2025annoation} reported that textual annotations appeared in 76\% of  the 1,888 charts they analyzed). Moreover, we recommend the following operational criterion: an annotation must have a bound target in the visualization~\cite{ren2017chartaccent} and should not be an entirely independent textual element.

Other textual forms are longer and carry a higher information load. For example, \textit{description} (N = 35) typically appears as one or more paragraphs that summarize the basic information and key insights of the visualization. 
\textit{Story/document} (N = \x{34}) denotes the longest and most narrative-driven textual form; it can contain rich content that is both data-driven and supplemented with extensive additional information.
Lastly, two special cases have also been identified: \textit{alternative text} (N = 6) embedded within the visualization to provide screen-reader access for visually impaired users (which is usually a special type of description), and \textit{video subtitles} (N = 1) displayed while a data video is playing.


\subsection{Modalities of Integration}

We found that the relationship between text and visualization is dynamic and exists on a fluid spectrum, defined by their relative dominance (as shown in \autoref{fig:manifest}).

At one end of the spectrum lies \textit{text in visualization} (N = 41). In this circumstance, the visualization is the dominant object, and text serves in a subordinate, supporting role. This is characteristic of small-scale text forms like labels and annotations, which are physically embedded within the chart's graphical marks and margins. In our collected papers, this generally corresponds to single-chart design studies. For example, Ren~\etal~\cite{ren2017chartaccent} proposed a design space for annotation and developed a system to support realizing annotations on individual visualizations (\autoref{fig:manifest} a). Similarly, some researchers proposed single-view systems, such as Kauer~\etal~\cite{kauer2025towards}, who developed an interactive map interface on which anyone can annotate their stories.
In addition, we also identified two special cases. One is labeled data physicalization (\autoref{fig:manifest} b), i.e., affixing paper labels to 3-D printed visualizations~\cite{sauve2022put}. The other is labeled VR visualization (\autoref{fig:manifest} c), where users inspect individual label information while exploring 3D scientific particle datasets in immersive environments~\cite{novotny2024evaluating}.

A more balanced, \textit{hand-in-hand} (N = \x{57}) relationship is typical for longer text such as descriptions and stories, where text and visualization are treated as equal partners. Representative examples include data-driven articles with side-by-side or staggered text–visual layouts (\eg \autoref{fig:manifest} d).
Studies have also examined slideshows for data presentation (\autoref{fig:manifest} e), multi-view systems containing multiple panels, including dashboards (\autoref{fig:manifest} f), and more deeply integrated forms such as infographics and data comics (\autoref{fig:manifest} g).


At the opposite end of the spectrum is \textit{visualization in text} (N = 12). In this case, the textual narrative—such as a report or document—is the primary medium. Visualizations are embedded as supporting evidence within the flow of the text, and their size and placement are more or less subordinate to the text. For example, Goffin~\etal~\cite{goffin2016exploratory} conducted an exploratory study examining how designers embed various micro-graphics into documents to enhance readability and expressiveness (\autoref{fig:manifest} h). Such designs have also been observed in scientific documents (\eg research papers), where researchers often add small visualizations into text to save space and facilitate quick comparison (\autoref{fig:manifest} i).

These findings remind us that text is not necessarily subordinate to visualization. In a considerable body of work, text and visualization already act as equal partners, or text even assumes a more dominant role, serving diverse design forms and communication scenarios.
Another key key insight is that these relationships are not fixed; they can be layered, combined, or switched dynamically. 
For example, a frequently observed nesting pattern, shown in \autoref{fig:manifest} j, is to employ a relatively balanced text–visual layout in a data-driven article, while each visualization itself is further reinforced by internal titles, annotations, and other micro-texts that elaborate the story.
Similarly, \autoref{fig:manifest} k presents a system for code-quality analysis that uses text to augment the presentation of analytical findings. It retains the conventional multi-panel structure of visual-analytics systems—dedicating separate panels to text and graphics—yet embeds small inline figures within the text panel to facilitate browsing and navigation.

Interestingly, we also identified three papers that support seamless transitions among multiple text–visual integration modes. For instance, Beck~\etal~\cite{beck2017word} implemented several assembly forms of text and image, allowing users to juxtapose a visualization with text or embed it inline (\autoref{fig:manifest} m).
Charagraph~\cite{masson2023charagraph} (\autoref{fig:manifest} l), similarly, enables users flip fluidly between text and chart—numbers become interactive graphics in one click, and graphics can be collapsed back into text—so data-heavy paragraphs are instantly explorable without ever leaving the sentence.

Although still limited in number, we believe such works carry significant weight, as they \textbf{fundamentally challenge the long-held view of text as a static, linear medium meant to be read sequentially and interpreted in isolation}. Instead, they reconceptualize text as a dynamic, interactive, and composable multimedia element. In this new paradigm, textual components can be directly manipulated, dynamically linked to visual entities, reorganized based on user interaction, and even serve as interactive controls that trigger updates in visual representations. This transformation elevates text from a passive explanatory supplement to an integral, active participant in the analytic discourse, enabling more fluid, non-linear, and user-driven exploration of data narratives. Such a transition mirrors broader movements in interactive media towards fluid documents and malleable content, positioning text not as an endpoint, but as a versatile interface to the underlying data and insights.

\subsection{Manifestations Across Research Areas}

By referring to the analysis framework of previous work~\cite{rahman2025annoation}, we coded the research areas of text as narrative in the context of data visualization. Note that different from the specific domain needs reported in \autoref{ssec:domain_needs}, this part pays more attention to which research directions and topics the various text–viz integrations are used in. For example, the medical and biology industries may both need to enhance narrative while analyzing data, so related studies can all be classified under the research direction of \textit{Exploratory Data Analysis} (EDA). The journalism and education industries may both need to tell stories with data, so such papers are placed in the broad category of \textit{Storytelling and Narrative Visualizations}.

From the statistical results, the number of papers related to \textit{Storytelling and Narrative Visualizations} is the highest (N = \x{30}). This is also quite intuitive, because effectively communicating insights to a broad audience often requires a compelling narrative structure, where text serves as a critical tool to guide interpretation, provide context, and emphasize key messages—functions that are essential in fields such as journalism, advertising, and public communication. For instance, tools like Idyll Studio~\cite{conlen2021idyll} and DataTales~\cite{sultanum2023datatales} focus on weaving data-driven narratives through tightly coupled text and visual elements.
Next come \textit{General Information Visualization Design} (N = 23), \textit{Exploratory Data Analysis} (N = 19), \textit{Interactive Document} (N = 14), \textit{Accessible Visualization} (N = 9), \textit{Tangible Visualization} (N = 2), and \textit{Collaborative Visualization} (N = 2).

Relating the previously analyzed text-visualization integration forms to these research areas reveals distinct patterns. For example, within the \textit{Storytelling and Narrative Visualizations} category, 19 out of 30 cases employ integrated forms such as data-driven articles, slideshows, and data videos—formats commonly consumed by the general public in mainstream scenarios. In contrast, the \textit{Exploratory Data Analysis} category is characterized by a continuation of the visual analytics paradigm, with 11 out of 19 cases proposing multi-view systems that incorporate text as narrative to enhance overall communication effectiveness and user experience. Furthermore, in the \textit{General Information Visualization Design} category—often evaluated through controlled experiments—14 out of 23 cases focus on single-chart or single-view systems. These findings suggest that the integration of text and visualization is not only a domain-specific need but also a cross-cutting concern that is highly relevant to a wide range of research directions.

%% file: Sections/06-how.tex
\section{How to Perform Text Design?}

Our analysis of the \textit{how} dimension has led us to group existing design techniques into five primary narrative tasks (T1-T5) and a set of sub-tasks.
\x{
The five main narrative tasks that text serves are: \textit{explain}, \textit{emphasize}, \textit{couple}, \textit{adapt}, and \textit{verify}. Among these, design techniques for the \textit{explain} task are primarily \textbf{content-oriented}, focusing on semantic framing, information organization, and contextual enrichment to support data interpretation.
Design techniques for the \textit{emphasize} and \textit{couple} tasks are largely \textbf{form-oriented}, involving visual appearance, layout, and interactive linkages to guide attention and bridge text-visual separation. While \textit{adapt} and \textit{verify} are less frequently addressed in current literature, we argue that they represent equally important—and increasingly relevant—opportunities. 
\textit{Adapt} tailors narratives across contexts and media, blending content restructuring with form redesign. \textit{Verify} ensures trust in AI-era narratives through content critique and visual confidence cues. Therefore, design techniques for these tasks are typically more complex and \textbf{hybrid in content and form} to enable seamless modality switching, user-controlled critique, and trustworthy narrative collaboration.
} See \autoref{tab:design} for the summarization in detail.

\input{Tables/design}

\subsection{T1: Explain}

The task \textit{explain} involves using text to facilitate the interpretation of insights from visualizations, thereby enhancing the audience's comprehension of the data. We identify four main strategies within this task.

\begin{figure*}[t]
 \centering
 \includegraphics[width=\textwidth]{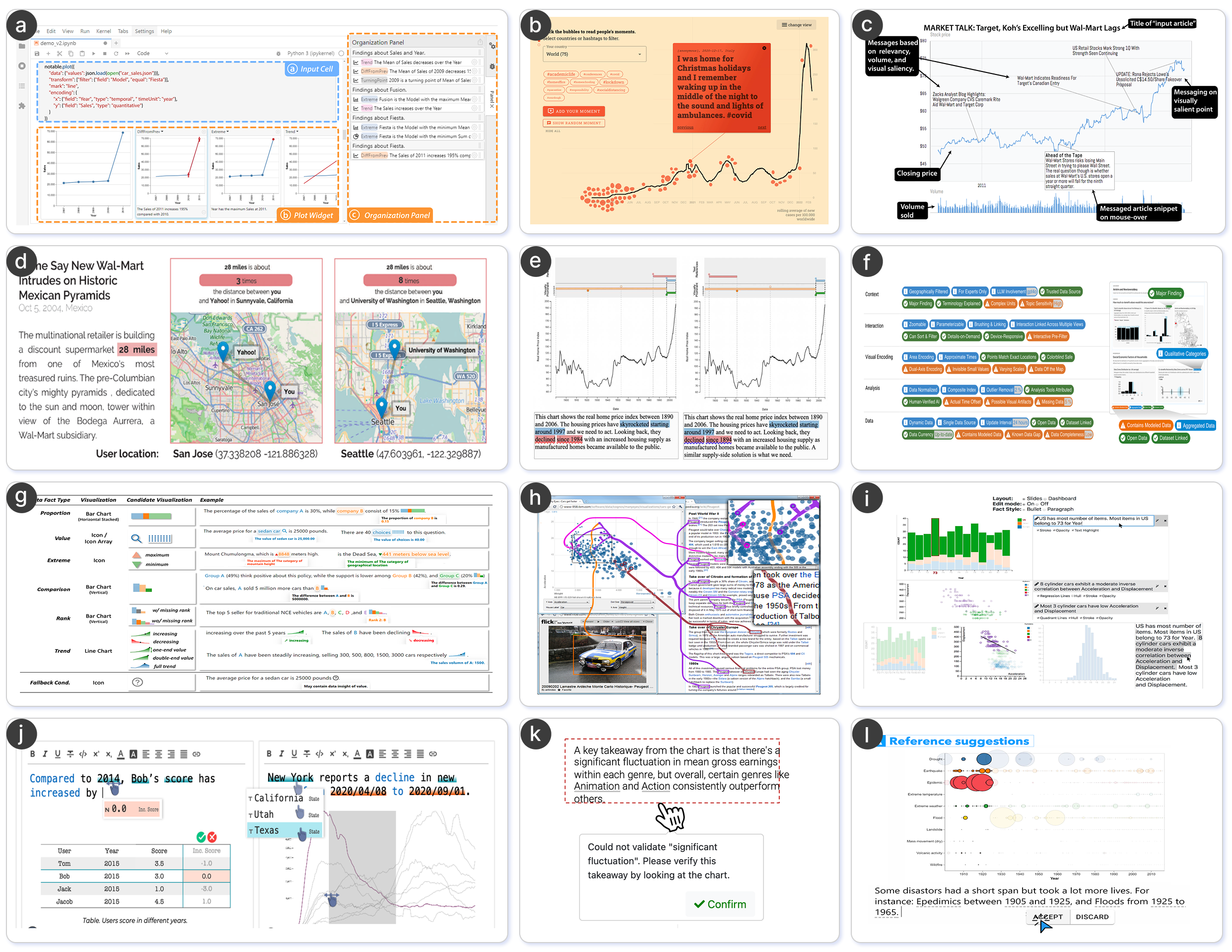}
 \caption{Examples of various design techniques: (a) Notable~\cite{li2023notable}, (b) coronaMoments~\cite{kauer2025towards}, (c) Contextifier~\cite{hullman2013contextifier}, (d) Personalized sentence~\cite{kim2016generating}, (e) EmphasisChecker~\cite{kim2023emphasischecker}, (f) Visualization badges~\cite{edelsbrunner2026visualization}, (g) GistVis~\cite{zou2025gistvis}, (h) Visual links~\cite{steinberger2011context}, (i) Voder~\cite{srinivasan2018augmenting}
 (j) Crossdata~\cite{chen2022crossdata}, (k) Pluto~\cite{srinivasan2025pluto}, (l) Kori~\cite{latif2021kori}.}
 \label{fig:examples}
 \Description{A 3x4 grid of twelve panels, labeled (a) through (l), showcasing diverse examples of integrating text within data visualizations. Refer to the main text of the paper for detailed descriptions of each example.}
 \vspace{1em}
\end{figure*}

\textbf{T1.1: Organize explanatory text with a narrative structure.} This strategy involves structuring the content to be explained through specific narrative logics. One solution is to transform data content into a hierarchical structure organized by themes and granularity. For instance, Notable~\cite{li2023notable} (\autoref{fig:examples} a) summarizes the analytical process into operational units and sub-units, which are then presented as textual bullet points in slides for communication. Beyond simple hierarchical relationships, other researchers have drawn inspiration from more advanced narrative logics in storytelling, such as the inverted pyramid, parallel structures, and the 4Ws (Who, What, When, Where) model, to organize language. For example, DataParticles~\cite{cao2023dataparticles} structures its descriptive narratives of visualizations using the inverted pyramid model.

\textbf{T1.2: Incorporate contextual information.} Explanatory power can be enhanced by integrating rich contextual information. Common methods include allowing users to manually add contextual annotations. For example, Kauer~\etal~\cite{kauer2025towards} (\autoref{fig:examples} b) developed an interactive map that enables users to collectively annotate their thoughts and experiences about COVID-19. Other approaches automatically incorporate relevant context and knowledge from external sources. For instance, ChartStory~\cite{zhao2021chartstory} adds hyperlinks to key concepts in the text, directing users to Wikipedia pages. Contextifier~\cite{hullman2013contextifier}, a system generating financial data visualizations (\autoref{fig:examples} c), extracts the day's latest financial news; important context influencing asset price movements is then integrated into the visualization as textual annotations. Another visualization-specific method for explanation involves disclosing the transparency of data processing and analysis. A common solution is to use tooltips or metadata text near visualizations to provide supplementary methodological details. For example, VIS Author Profiles~\cite{latif2018vis} shows an interface for analyzing researchers' publication activity, which includes a small icon next to metrics like ``co-author'' that reveals explanatory details upon hover. Recently, Edelsbrunner~\etal~\cite{edelsbrunner2026visualization}, through iteractive design process, proposed a design toolkit of standardized badges (\autoref{fig:examples} f) that can be added to any visualization to enhance the analysis and design transparency.


\textbf{T1.3 Manage perspectives and insight content.} This strategy determines the angle and content from which an explanation is approached, which is a key communication technique. For presenting meaningful insights, one approach allows users to configure the diversity or feature importance of the insights they wish to receive, thereby extracting as many useful insights as possible from a given dataset. Lundgard~\etal~\cite{lundgard2021accessible}, for example, proposed a four-level model, recommending that narration can begin with elemental and encoded information, followed by statistical and relational, perceptual and cognitive, and finally, contextual and domain-specific information. Other solutions include switching the frames of data narratives or creating user-perspective personal data descriptions to ensure the communicated content is familiar, relevant, and valuable to the audience. For example, Kim~\etal~\cite{kim2016generating} transformed abstract geographic distance into personalized ``perspective sentences'' (\autoref{fig:examples} d).

\textbf{T1.4: Enrich communication styles.} Finally, explanation is also concerned with adapting communication styles. For instance, the style can be either concise or elaborate, corresponding to the level of detail in the explanatory text. For example, when generating accessible text descriptions for visualizations, Mcnutt~\etal~\cite{mcnutt2025accessible} offered two modes, succinct and verbose, allowing users to choose their preferred level of detail. 
A recent work powered by LLMs~\cite{yu2024towards} allows users to choose their desired length (\eg short, medium, long) before generating chart summarization.
Second, the style can be scientific and abstract (e.g., for domain experts) or intuitive and approachable (e.g., for the general public)~\cite{lai2020automatic}. For instance, analogical expressions can be used to transform abstract and complex data descriptions into more straightforward forms~\cite{riederer2018put}.
Another identified approach is to configure the sentiment and naturalness of the language~\cite{yu2024towards}. Explanations can range from calm and direct to more emotional and expressive.

\subsection{T2: Emphasize}

The task \textit{emphasize} involves using textual properties and annotations to direct the audience's attention to specific data entities, insights, or issues within a visualization, ensuring that the most critical information is prominently noticed.

\textbf{T2.1: Highlight key entities.} This is a frequently addressed design task, with two primary solutions. The first solution involves the use of typographic attributes for highlighting, such as font color, font weight, and font background (\eg \autoref{fig:examples} a, d, e). This method is extremely common in text design, and consequently, a large number of papers employ it. It is worth noting, however, that the majority of these papers apply specific highlighting techniques without delving into the underlying rationale. A notable exception is the work by Choudhry et al.~\cite{choudhry2020once}, which meticulously catalogued the specific highlighting effects needed before implementing them in a visualization system. Another significant contribution is from Strobelt et al.~\cite{strobelt2015guidelines}, who conducted user experiments to compare nine different text highlighting techniques, providing a systematic empirical evaluation.
The second solution is to add \textit{observational} textual annotations near key data insights. In contrast to the \textit{contextual} annotations discussed in T1 (which typically incorporate knowledge external to the data), observational annotations are derived from the data itself and primarily serve to emphasize important data values, categories, or patterns.

\textbf{T2.2: Enhance readability.} Emphasis and the reduction of visual clutter are two sides of the same coin. Highlighted textual elements can only be effective when all text elements are arranged appropriately and serve distinct purposes. Therefore, we have also identified several attempts to enhance text legibility, such as employing critical print sizes (\eg ~\cite{novotny2024evaluating}), and placing text near points of interest (\eg \cite{lai2020automatic}).

\textbf{T2.3: Augment semantics.} Beyond purely visual highlighting, designers can augment emphasized elements with additional semantic information. For example, we observed solutions that use graphics, such as color and shape, to convey specific meanings. Kim et al.~\cite{kim2023emphasischecker} used a red squiggly underline to indicate a mismatch between the text description and the visualization (\autoref{fig:examples} e). Similarly, CrossData~\cite{chen2022crossdata} employed wavy shapes to emphasize errors or warnings in text.

\subsection{T3: Couple}

The third task, \textit{couple}, aims to establish and enhance the referential relationships between textual elements and their corresponding components in a data visualization. This task is crucial for creating a tightly coherent and fluid data-driven narrative.

\textbf{\x{T3.1: Inline enhancement.}} This strategy involves embedding word-scale graphics directly into the textual flow. For example, an early collaborative visualization by Willett~\etal~\cite{willett2007scented} embeds small line charts to summarize the editing activities of different collaborative users. Goffin et al.~\cite{goffin2014exploring,goffin2016exploratory,goffin2020interaction} conducted a series of studies on word-scale graphics, such as observing how designers produce word-scale graphics to enhance the communication of data-rich documents based on the data facts they present (e.g., proportion, rank). More recently, GistVis~\cite{zou2025gistvis} proposed a method to automatically extract key data entities from text and generate corresponding micro-visualizations alongside them (\autoref{fig:examples} g).

\textbf{T3.2: Establish text-visualization reference.} This strategy focuses on creating explicit binding relationships between textual and visual elements, aiding users in understanding the origins of conclusions and judgments made in the text. A substantial body of research addresses this through interactive linking. For instance, in Voder~\cite{srinivasan2018augmenting} (\autoref{fig:examples} i), when users hover certain textual entities, the corresponding part of the visualization is highlighted. The system Kori~\cite{latif2021kori} supports three types of text-visual bindings: point-level matching, interval-level matching, and reference grouping. We also identified a case that use visual lines to directly link text and visualization elements (\autoref{fig:examples} h).
Regarding the specific cues used to signal these linkages, we observed that \textbf{fading} is a predominant technique (as shown in (\autoref{fig:examples} i). 
\textbf{Identical coloring} is also an common cue for linking; when the color of a visualization element matches that of a specific part of the text, users are more likely to perceive them as related~\cite{goffin2016exploratory,kong2014extracting}.

\textbf{T3.3: Mutual translation.}
This strategy represents the most advanced paradigm for coupling text and visualization, transforming their relationship from one of static reference to dynamic, bidirectional partnership. Here, text and visualization are co-generated and mutually editable, forming an integrated environment for sensemaking. This paradigm is exemplified by systems like CrossData~\cite{chen2022crossdata} (\autoref{fig:examples} j), DataWeaver~\cite{fu2025dataweaver} and Pluto~\cite{srinivasan2025pluto}, where edits to the textual narrative (e.g., modifying a data fact or insight) automatically propagate to update the visualization, and conversely, interactions with the visualization (e.g., filtering or highlighting) are instantly reflected in the textual description. 
The significance of this paradigm extends beyond mere technical coupling; it fundamentally transforms the data storytelling process by enabling a continuous translation between visual patterns and textual statements, thereby merging the traditionally separate modes of exploratory analysis and explanatory communication into a single, fluid practice.

\subsection{T4: Adapt}

The task \textit{adapt} focuses on optimizing the presentation of the text-visualization narrative to fit specific media formats, user preferences, or communication scenarios. It ensures that the integrated representation is appropriate for its final medium of dissemination.

\textbf{T4.1: Support multiple text-visualization integration modes.} This strategy provides flexibility in the spatial arrangement and joint presentation of text and visuals. For instance, systems like DataParticles~\cite{cao2023dataparticles} and TimelineCurator~\cite{fuld a2015timelinecurator} support multiple output formats, such as exporting the system’s web interface directly or offering text-visualization integration modes better adapted to mobile devices.
A few studies~\cite{masson2023charagraph,beck2017word} goes even further. These works systematically catalog integration modes along a spectrum ranging from text-in-chart to chart-in-text, as introduced in \autoref{sec:space}, thereby providing designers with dynamic switching capabilities among these options.

\textbf{T4.2: Support format transfer.} This strategy involves transforming the narrative from one format to another, effectively ``repurposing'' content for a different medium or context. This is crucial for streamlining the data communication workflow. Examples include automatically converting an analytical Jupyter notebook into a set of presentation slides~\cite{li2023notable,wang2023slide4n,zheng2022telling} (\autoref{fig:examples} a) \x{and transforming digital design into data physicalization~\cite{moured2024chart4blind}}.

\x{
\textbf{T4.3: Support different communication scenarios.} This strategy primarily addresses diverse communication needs. For example, Arunkumar~\etal~\cite{arunkumar2025lost} considered the visual information processing patterns of multilingual communities and proposed a series of design suggestions to accommodate cultural and language preferences. Meanwhile, Zheng~\etal~\cite{zheng2022telling}, when generating text-visualization-integrated slides, took into account the needs of both technical and non-technical audiences.
}


\subsection{T5: Verify}

\textit{Verify} addresses the critical aspect of trust and accuracy in data-driven narratives. As automated text generation becomes prevalent, especially with LLMs, this task provides mechanisms for users to assess, critique, and establish confidence in the presented narrative.

\textbf{T5.1: Support manual critique and refinement.} This strategy allows users to directly interrogate and edit the narrative, treating it as a malleable object rather than a static assertion. For example, \autoref{fig:examples} i provides an edit mode, enabling users to freely modify the chart's descriptions if they find them inaccurate or not satisfying. This active participation fosters a more critical engagement with the textual narrative.
Or, some systems enable a verification widget in text that allows users to judge whether to accept or reject textual narratives generated by LLMs (\eg \autoref{fig:examples} k, l)~\cite{srinivasan2025pluto,latif2021kori}. These cues make the system's uncertainty or the need for human validation explicit, promoting transparency and preventing automated overconfidence.

\subsection{\x{Summary}}
In summary, our analysis categorizing text design techniques from the literature into five narrative tasks reveals a clearer research landscape. The most extensive body of work is concentrated on leveraging the function of text to \textit{explain} and \textit{emphasize}, which form the foundational core of textual narration. Meanwhile, \x{a substantial number of studies revolve around \textit{couple}, primarily employing various text-visualization binding techniques to enhance cross-modal connections. This reflects the visualization community's progress in interactive design and system development.} In contrast, the tasks of \textit{adapt} and \textit{verify} represent emerging and underexplored frontiers, often not being explicitly addressed as primary design goals. At the level of specific strategies, the visualization community has produced a rich suite of solutions for techniques such as visual highlighting (T2.1), embedding word-scale graphics (T3.1), and establishing text-visualization reference (T3.2). Other strategies, particularly those within the understudied tasks, offer significant opportunities for future research.

%% file: Tables/design.tex

\makeatletter
\def\arraystretch{1.4}      
\makeatother

\begin{table*}[htbp]
    \fontsize{8}{8.7}\selectfont
    \centering
    \begin{tabular}{p{1.5cm}|p{3cm}|p{5.3cm}|p{6.5cm}}
    \toprule
    \rowcolor[HTML]{E4E9FF} 
    \textbf{Narrative tasks} & \textbf{Sub-tasks} & \textbf{Design techniques} & \textbf{Literature}\\ \hline 

    \multirow{16}{*}{\colorbox{c1!50}{\textbf{Explain}}} 
    {} & \multirow{4}{=}{T1.1 Organize explanatory text with a narrative structure} & {\x{Hierarchical structure (\eg lists, bullet points)}} & {\cite{choudhry2020once,srinivasan2018augmenting,wang2023slide4n,li2023notable,zheng2022telling,lecardonnel2025genqa,sultanum2024instruction,mcnutt2025accessible,zhao2021chartstory,pan2023graphdescriptor}} \\ 
    {} & {} & {Reverse pyramid structure} & {\cite{cao2023dataparticles,yang2021design}}  \\ 
    {} & {} & {Parallel structure} & {\cite{cao2023dataparticles}}  \\ 
    {} & {} & {4Ws (Who, What, When, Where)} & {\cite{metoyer2018coupling}}  \\ \cline{2-4}
    {} & \multirow{6}{=}{T1.2 Incorporate contextual information} & {Manual addition \x{of} contextual annotations} & {\cite{satyanarayan2014authoring,bryan2016temporal,sauve2022put,kauer2025towards,gao2014newsviews,sultanum2024instruction,rahman2024qualitative,wang2020data,elias2012annotating,heer2007voyagers}} \\
    {} & {} & {Hyperlink to external knowledge bases} & {\cite{zhao2021chartstory,latif2021deeper}}  \\
    {} & {} & {Extract and present context (\eg from news)} & {\cite{hullman2013contextifier}}  \\ 
    {} & {} &  {Tooltips showing pre-set definition/explanation} & {\cite{latif2018vis,sultanum2018doccurate,zou2025gistvis,latif2021deeper,mumtaz2019exploranative}} \\
    {} & {} & {Plain text showing data sources/methodology} & {\cite{sultanum2024instruction,burns2022invisible,elias2012annotating,wang2020data,cao2023dataparticles,zheng2022telling,latif2021deeper,mumtaz2019exploranative}}  \\ 
    {} & {} & {\x{Standardized badges}} & {\x{\cite{edelsbrunner2026visualization}}}  \\ 
    \cline{2-4}
    {} & \multirow{3}{=}{T1.3 Manage perspectives and insight content} & {Semantic levels (\eg four-level model)} & {\cite{lundgard2021accessible,stokes2022more,stokes2022striking,moured2024chart4blind,fan2024understanding,li2024altgeoviz,bromley2024dash,srinivasan2025pluto,mcnutt2025accessible,xu2024graphs,jung2021communicating}}   \\ 
    {} & {} & {Switch frames} & {\cite{kong2018frames,stokes2023role,kong2019trust}}  \\
    {} & {} & {Employ a personalized perspective} & {\cite{wang2023making,kim2016generating,riederer2018put,liem2020structure}}  \\ \cline{2-4}
    {} & \multirow{3}{=}{T1.4 Enrich communication styles} & {Manage text length/detail level} & {\cite{yu2024towards,mcnutt2025accessible,conlen2021idyll,seo2024maidr,sultanum2024instruction,liu2023autotitle,fan2024understanding,alam2023seechart,jung2021communicating,hoffswell2018augmenting}} \\ 
    {} & {} & {Analogical expression} & {\cite{kim2016generating,riederer2018put,conlen2021idyll}}  \\  
    {} & {} & {Adjust naturalness/informativeness} & {\cite{yu2024towards,liu2023autotitle,zhao2021chartstory}}   \\
    \hline

    \multirow{6}{*}{\colorbox{c2!50}{\textbf{Emphasize}}} & \multirow{4}{=}{T2.1 Highlight key entities} & {Modify typographic attributes (\eg font color, weight, background, underline)} & {\x{\cite{brandes2011asymmetric,kong2014extracting,badam2018elastic,zhi2019linking,choudhry2020once,mumtaz2019exploranative,latif2018vis,steinberger2011context,strobelt2015guidelines,zheng2022evaluating,kim2023emphasischecker,metoyer2018coupling,lai2020automatic,chen2022crossdata,masson2023charagraph, parnow2015micro, kim2016generating, latif2018exploring,latif2021kori,wang2022towards,dragicevic2019increasing,chen2010click2annotate,sultanum2024instruction,kandogan2012just,fulda2015timelinecurator,sultanum2018more,sultanum2018doccurate,barral2020understanding,sultanum2023datatales,fu2025dataweaver,gao2014newsviews,lecardonnel2025genqa,bromley2024dash,lalle2021gaze,cai2024linking,zou2025gistvis,xu2024graphs,shen2023data,zhao2021chartstory,pan2023graphdescriptor,li2023notable,cao2023dataparticles,ajani2021declutter,wang2020data,latif2021deeper}}}    \\ 
    {} & {} & {Observational annotations} & {\cite{eccles2008stories,kong2012graphical,kandogan2012just,gao2014newsviews,ren2017chartaccent,bryan2016temporal,lai2020automatic,latif2021deeper,rahman2024qualitative,chen2023calliope,li2024viscollage,arunkumar2025lost,stokes2022striking,alam2023seechart,stokes2022more,ajani2021declutter}}   \\ \cline{2-4}

    {} & \multirow{2}{=}{T2.2 Ensure readability} &  {\x{Text legibility (\eg critical print size)}} & {\x{\cite{sultanum2024instruction,novotny2024evaluating,moured2024chart4blind}}}  \\
    {} & {} & {Place text near the point of interest \x{(POI)}} & {\cite{lai2020automatic,bryan2016temporal,steinberger2011context,kong2012graphical,kandogan2012just,li2024viscollage,gao2014newsviews,chen2023calliope}}  \\ \cline{2-4}
    
    {} & \multirow{1}{*}{T2.3 Augment semantics} & {Semantically-resonate color/shape} & {\cite{chen2022crossdata,kim2023emphasischecker,sultanum2018more}}   \\  \cline{2-4}
    \hline

    \multirow{7}{*}{\colorbox{c3!50}{\textbf{Couple}}} & 
    \multirow{2}{=}{T3.1 \x{Inline enhancement}} & {Embed word-scale graphics} & {\cite{willett2007scented,lee2010sparkclouds,brandes2011asymmetric,brandes2013gestaltlines,parnow2015micro,goffin2016exploratory,beck2017word,latif2018exploring,choudhry2020once,mumtaz2019exploranative,heer2009sizing,beck2016expert,latif2018vis,sultanum2024instruction,zou2025gistvis,cai2024linking,li2023notable,goffin2020interaction,goffin2014exploring}}  \\  \cline{2-4}

    {} & \multirow{3}{=}{T3.2 Establish text-visualization reference} & {Bounded effects (\eg fade, color, animation)} & {\x{\cite{kong2014extracting,parnow2015micro,badam2018elastic,srinivasan2018augmenting,latif2018exploring,metoyer2018coupling,zhi2019linking,zheng2022evaluating,masson2023charagraph,kim2023emphasischecker,sultanum2021leveraging,kwon2014visjockey,goffin2020interaction,chen2022crossdata,choudhry2020once,latif2021kori,mumtaz2019exploranative,lai2020automatic,beck2017word,xu2024graphs,latif2021deeper,willett2007scented,elias2012annotating,fulda2015timelinecurator,mckenna2017visual,sultanum2018more,barral2020understanding,conlen2021idyll,sultanum2023datatales,shen2023data,fu2025dataweaver,lecardonnel2025genqa,bromley2024dash,lalle2021gaze,cai2024linking,srinivasan2025pluto,pan2023graphdescriptor,cao2023dataparticles,zheng2022telling,sultanum2018doccurate,eccles2008stories,heer2007voyagers}}}   \\       
    {} & {} & {Visual lines} & {\cite{steinberger2011context}}  \\ \cline{2-4}    

    {} & \multirow{1}{=}{T3.3 Mutual translation} & {Bi-directional construction} & {\cite{dragicevic2019increasing,masson2023charagraph,fu2025dataweaver,srinivasan2025pluto,chen2022crossdata,mumtaz2019exploranative,bromley2024dash}}   \\  

    \hline

    \multirow{6}{*}{\colorbox{c4!50}{\textbf{Adapt}}} & {T4.1 Support multiple text-viz integration formats} & {Provide multiple integration choices} & {\cite{srinivasan2018augmenting,sultanum2021leveraging,masson2023charagraph,fulda2015timelinecurator,cao2023dataparticles,conlen2021idyll,latif2021kori,goffin2020interaction,wang2019comparing,mckenna2017visual}}   \\ \cline{2-4}
    {} & \multirow{3}{=}{T4.2 Support format transfer} & {Document to visualization} & {\cite{masson2023charagraph,badam2018elastic,chen2022crossdata}}   \\ 
    {} & {} & {Analysis notebook to presentation slides} & {\cite{li2023notable,wang2023slide4n,zheng2022telling}}  \\ 
    {} & {} & {\x{Digital to physical visualization}}  & {\x{\cite{moured2024chart4blind}}} \\ 
    \cline{2-4}
    {} & \multirow{2}{=}{\x{T4.3 Support different communication scenarios}} & {\x{Design for cultural and language preference}} & {\x{\cite{arunkumar2025lost}}}   \\ 
    {} & {} & {\x{Design for technical/non-technical audience}} & {\x{\cite{zheng2022telling}}}   \\
    \hline

    \multirow{2}{=}{\colorbox{c5!50}{\textbf{Verify}}} & \multirow{2}{=}{T5.1 Support manual critique and refinement} & {Enable manual edit and correction} & {\cite{srinivasan2018augmenting,satyanarayan2014authoring,xu2024graphs,chen2010touch2annotate,chen2010click2annotate,latif2021kori,zhao2021chartstory,lecardonnel2025genqa,sultanum2023datatales,conlen2021idyll,choudhry2020once}}   \\ 
    {} & {} & {Confirmation widgets} & {\cite{latif2021kori,srinivasan2025pluto}}   \\ 
    
    \bottomrule
    \end{tabular}
    \caption{A summary of existing literature that offers design implications for the identified narrative tasks.}
    \Description{A summary of existing literature that offers design implications for the identified text design tasks, organized according to the five identified narratives tasks: Explain, Emphasize, Couple, Adapt, and Verify. For each main task, it lists specific sub-tasks, corresponding design techniques, and supporting literature citations. You can interact with the table directly.}
    \label{tab:design}
    \vspace{0em}
\end{table*}

%% file: Sections/07-discussion.tex
\section{Discussion}

By reviewing 98 papers that contribute insights about text as narrative in the context of data visualization, this survey provides a unique, complementary perspective to previous surveys on text mining and natural language interfaces, thereby helping build a more comprehensive picture of text research in visualization.
Below, we discuss research gaps and opportunities we observe along the why-what-how dimensions.

\subsection{Research Gaps and Opportunities}

\subsubsection{About Why: The Need for Deeper and Broader Empirical Evidence}

The \textit{why} dimension is established but requires maturation. While the high-level value of text is well-acknowledged, the empirical foundation needs strengthening.

\textbf{Re-examining and reconciling inconsistent results:} Our synthesis reveals inconsistencies in some empirical findings (see \autoref{tab:empirical}). While most studies affirm the positive role of text, some report neutral or negative effects on metrics like efficiency and attitude change. This necessitates further research to re-test and reconcile these results. Future work can also investigate more variables (e.g., data complexity, user expertise, task type) that explain when and why text is beneficial, ineffective, or disruptive.

\textbf{More nuanced empirical studies:} Future research may move beyond general questions of ``is text useful?'' to investigate under what conditions, for which tasks, and for which audiences specific textual techniques are most effective. For example, \x{an interesting finding from Lall{\'e}~\etal~\cite{lalle2021gaze} is that linked text-visual highlighting significantly improved understanding among low-literacy users, but provided insignificant benefit to high-literacy users. Similarly, we can pose many more questions like:}
Is an analogical explanation more effective for novices than experts? \x{How should textual framing and detail be adapted to different communicative contexts?}

\textbf{Beyond Classic Metrics}: Beyond classic metrics such as comprehension, memorability, and subjective experience, \x{our analysis (as summarized in \autoref{tab:empirical}) suggests that metrics such as trust, attitude change, data prediction, and exploration behaviors remain under-explored. We argue that these metrics are particularly valuable in specific application scenarios. For instance, attitude change is central to persuasive visualizations~\cite{pandey2014persuasive}, and text, as a powerful rhetorical tool throughout history, can be pivotal in shaping these outcomes~\cite{lin2025unveiling,lan2022negative}. The ability of prediction is crucial in scenarios involving trend analysis and decision-making under uncertainty~\cite{stokes2023role}. Perhaps most pressingly, trust has become a paramount issue in the age of AI-generated content.
}. As argued by some researchers~\cite{kong2018frames,zheng2022evaluating}, designing text to combat misinformation of data is particularly critical and emerging. \x{Investigating how text design and text-visualization integration shape the perceived credibility of data content represents a timely research direction.}

\subsubsection{About What: Expanding the Narrative Forms and Mediums}

\x{Although text is prevalently used in visualizations, its integration often remains constrained by conventions inherited from the printing era. We advocate for a re-imagination of text's role—exploring it as a dynamic and co-equal partner to visuals, particularly in emerging mediums where these traditional constraints can be transcended.}

\textbf{Beyond the screen:} Current research focuses overwhelmingly on desktop and mobile-screen-based visualizations. \x{To advance the \textit{what} dimension, we may look beyond these screens toward a paradigm of environmental computing.} This entails exploring text design for emerging mediums such as \textbf{data physicalization}~\cite{sauve2022put} and \textbf{immersive environments}~\cite{novotny2024evaluating}. \x{In this paradigm, text is a dynamic entity embedded in the user's physical or immersive space, shifting the experience of data narrative from viewing to inhabiting. These contexts also raise interesting questions such as:} How does textual annotation function on a 3D-printed model, and what new narrative forms become possible within a virtual space?

\textbf{Fluid text-vis relationships:} The traditional taxonomy of text-in-vis, vis-in-text, and hand-in-hand remains useful but static. \x{As shown in \autoref{fig:manifest}, we have already identified several papers that explore} more fluid and switchable relationships between text and visualization (\eg \cite{chen2022crossdata,masson2023charagraph}), \x{allowing designers to actively orchestrate the narrative by transitioning between different integration modes. 
In these papers,} text is not a static object; it can serve as an interactive and bidirectional gateway—clicking on a sentence can filter a chart, \x{and statistical claims can be made traceable to their source data or even recalculated in the document~\cite{dragicevic2019increasing}.
This pursuit resonates deeply with a broader \textbf{critique of rigid interfaces}, holding that overly formal systems stifle the exploratory thinking essential for deep understanding~\cite{shipman1999formality}. This critique once inspired the vision of \textbf{explorable explanations}~\cite{victor}, which stands in opposition to static, read-only documents, and today, it is becoming more technically feasible at scale. Looking forward, we envision that this research trajectory will further intersect with advances in natural language interfaces (i.e., the thread of \textit{text as interaction}), pushing explorable explanations beyond a reliance on predefined widgets and pre-programmed answers toward a future of highly reactive, personalized, and generative interaction.}

\subsubsection{About How: Systematizing and Evaluating Design Techniques}

\x{Below, we discuss research gaps corresponding to each of the identified narrative tasks.}

\textbf{T1: Explain}
While technical methods for summarizing data insights are abundant, the \textit{design} of explanations is less systematic. For each sub-task (e.g., using narrative structures, managing perspective), we found only limited papers offering concrete and practical design ideas. 
Future work needs to build a more comprehensive and generalizable repertoire of techniques. For instance, while Kim~\etal~\cite{kim2016generating} explored spatial analogies, their applicability to other data types is unknown. Also, it remains unclear which communication styles are best suited for different scenarios.
This gap in design knowledge, furthermore, presents a challenge when considering the use of LLMs to automate the \textit{Explain} task. Although LLMs show a remarkable capability to generate fluent explanatory text, determining how to guide them to produce outputs that are accurate, clear, and engaging for a specific context remains an open problem. 
Industrial practices have begun to confront this challenge. For instance, the AntV T8~\cite{t8} text visualization engine defines a structured narrative-text-schema—a format readily generated by LLMs—to transform data insights into narrative texts that incorporate highlighted key phrases and inline micro-charts. This exemplifies the approach of using an intermediate schema to structure and thereby control the final narrative output of LLMs.
Moreover, currently, explaining data analysis methodology often remains primitive, typically relying on static plain text and tooltips. Future research could explore more novel forms to convey transparency. A promising direction is exemplified by a recent work~\cite{edelsbrunner2026visualization}, which introduces standardized visualization badges to encode textual metadata about the data analysis process. This represents a move towards more structured and generalizable forms of explanation that go beyond plain text. 


\textbf{T2: Emphasize}
Regarding this task, a significant gap exists between practice and evidence. Although highlighting is widely used, we found only one systematic evaluation of text highlighting techniques~\cite{strobelt2015guidelines}. \x{While this work provides valuable insights, its focus on lookup tasks in textual documents does not fully address the complex needs of visual data analysis and communication. Consequently,} findings from this study (e.g., the efficacy of font size for highlighting) are not consistently applied in practice. \x{In many of the papers we surveyed, textual highlighting techniques (e.g., using boldface or background colors) were deployed based on convention rather than an evidence-based understanding of their effectiveness for specific analytical or narrative purposes.}
Furthermore, research on using design to create semantic resonance (T2.3) is nascent. There is a need to catalog common data semantics (e.g., urgency, importance, errors) and establish conventions for mapping them to visual/kinetic effects.

\textbf{T3: Couple}
\x{Word-scale graphics (T3.1) is a classic approach to coupling text and visualization. However, empirical evaluations of such designs are scarce compared to their numerous applications. As one example, Goffin et al. \cite{goffin2014exploring} compared several layouts for word-scale graphics to determine optimal placement and sizing. Yet, the lack of a baseline comparison with conventional charts makes it difficult to demonstrate the technique's incremental value, and} their study was limited to basic chart types (e.g., bar and line). Future guidelines should be extended to more complex charts (e.g., scatterplots, which suffer more from information loss when miniaturized) and incorporate rigorous baselines.
\x{Using interactive effects to link text and visualization is another major direction. Here,} the most significant gap and opportunity lie in bidirectional manipulation (T3.3). As discussed above, systems that enable true co-construction of text and visualization represent a paradigm shift, merging exploratory analysis and explanatory narrative into a fluid practice.
\x{Additionally, the powerful language understanding capabilities of LLMs can function as a semantic binding engine, enabling more precise, flexible bidirectional manipulation. This advancement would move coupling beyond simple data-point highlighting to the real-time interpretation of complex clauses and nuanced statements.}

\textbf{T4: Adapt}
This task is overall understudied, and there is a scarcity of literature on supporting multiple integration modes or genre transfers. \x{Besides, existing work primarily focuses on structurally clear data, such as converting the linear analysis and visualization process from a computational notebook into a linear slideshow~\cite{zheng2022telling,wang2023slide4n}. Given that LLMs have already demonstrated strong performance in such structured cross-modal tasks (e.g., extracting structured information from documents to create slides), the future research focus may shift toward more complex, unstructured data. Key challenges include how to} enable transformations that bridge varying information densities, \x{such as condensing a detailed data-driven article into a concise report without losing core insights, and how to achieve conversions involving non-linear or partly-linear forms, such as adapting data comics into other narrative genres.}
This is crucial for supporting data communication across different platforms and scenarios.

\textbf{T5: Verify}
This is a critical yet almost entirely unexplored frontier. In the age of AI-generated content, \x{verification becomes paramount yet profoundly challenging. LLMs can produce convincingly fluent but potentially misleading narratives, creating the narrative hallucination problem — where generated text appears authoritative but misrepresents the underlying data.} Future research is suggested to investigate techniques for supporting manual critique and refinement (T5.1) of data narratives and providing explicit verification cues (T5.2) to help users assess the validity of LLM-generated textual elements, such as titles, descriptions, and annotations. This is essential for developing responsible and trustworthy AI-assisted visualization systems.

\subsection{Limitations}

Our survey also has several limitations that point to directions for future research. First, while we aimed for a comprehensive review, our literature collection may have missed relevant work in venues that fall outside our search scope, so that we do not claim our corpus is exhaustive. Besides, the field is evolving rapidly with the rise of AI. Our taxonomy provides a foundation for understanding this evolution, but new narrative forms and techniques enabled by LLMs will likely necessitate future expansions and refinements of this framework.

%% file: Sections/08-conclusion.tex
\section{Conclusion}

This survey has systematically charted the role of text as a narrative medium in data visualization, moving beyond its conventional perception as a secondary element or one that is always submissive to graphics. Through a holistic why-what-how lens, we synthesize the field to establish a foundational understanding of textual narratives.

In terms of \textit{why}, we consolidate the empirical evidence and arguments for text's value in enhancing factors such as comprehension, engagement, and memorability, and in fulfilling critical domain-specific needs. In terms of \textit{what}, we categorize diverse textual forms and their integration modalities with visualization, revealing the varied and powerful roles text can play in data communication. Finally, our framework for \textit{how}—encompassing the tasks of explaining, emphasizing, coupling, adapting, and verifying—structures the design space, cataloging existing practices and exposing rich opportunities for future research.